\documentclass{IEEEoj}
\usepackage{cite}
\usepackage{amsmath,amssymb,amsfonts}
\usepackage{algorithmic}
\usepackage{graphicx,color}
\usepackage{textcomp}

\usepackage{tabularx}
\usepackage{bm}
\usepackage{braket}
\usepackage[caption=false,font=normalsize,labelfont=sf,textfont=sf]{subfig}
\usepackage{booktabs, makecell}
\usepackage{hyperref}
\usepackage{xurl}
\usepackage{xcolor}
\usepackage{balance}
\usepackage{multirow}

\def\BibTeX{{\rm B\kern-.05em{\sc i\kern-.025em b}\kern-.08em
    T\kern-.1667em\lower.7ex\hbox{E}\kern-.125emX}}
\AtBeginDocument{\definecolor{ojcolor}{cmyk}{0.93,0.59,0.15,0.02}}
\def\OJlogo{\vspace{-4pt}\hskip-4pt\includegraphics[height=18pt]{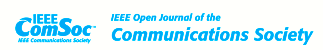}}

\newtheorem{remark}{Remark}

\begin{document}

\title{Quantum Annealing for Resource Allocation in Next-Generation Wireless Communication Networks: A QUBO Modeling Guide}

\author{NGUYEN XUAN TUNG\authorrefmark{1}, LE TUNG GIANG\authorrefmark{2}, MAI DINH CONG\authorrefmark{2}, \\PHAM DANG ANH DUC\authorrefmark{1}, NGUYEN HOANG VIET\authorrefmark{3}, \\TRINH VAN CHIEN\authorrefmark{3}, MEMBER, IEEE, AND LAJOS HANZO\authorrefmark{4}, LIFE FELLOW, IEEE}

\affil{Faculty of Interdisciplinary Digital Technology, PHENIKAA University, Yen Nghia, Ha Dong, Hanoi 12116, Viet Nam (e-mail: tung.nguyenxuan@phenikaa-uni.edu.vn, duc.phamdanganh@phenikaa-uni.edu.vn)}
\affil{School of Computer Science and Engineering, Pusan National University, Republic of Korea (e-mail: giang.lt2399144@pusan.ac.kr, cong.md@pusan.ac.kr)}
\affil{School of Information and Communications Technology (SoICT), Hanoi University of Science and Technology (HUST), Vietnam (e-mail:  nguyenhoangviet.hsgs@gmail.com, chientv@soict.hust.edu.vn)}
\affil{Electronics and Computer Science, University of Southampton, Southampton SO17 1BJ, U.K. (e-mail: lh@ecs.soton.ac.uk)}


\begin{abstract}
Resource allocation in next-generation (NG) wireless communication networks, including uncrewed aerial vehicle (UAV)-assisted networks, and space-air-ground integrated networks (SAGINs),
often involves discrete-variable combinatorial formulations associated with tightly coupled constraints, leading to high complexity and poor scalability under conventional solvers. Quantum annealing (QA) offers a physics-inspired approach for such subproblems by mapping them onto an Ising Hamiltonian, which is algebraically equivalent to a Quadratic Unconstrained Binary Optimization (QUBO) model executable on quantum annealers or hybrid solvers. However, their practical employment in wireless systems critically depends on the specific problem formulation, on the particular variable encoding, constraint handling, solver selection, hardware embedding, and fair benchmarking. This paper develops wireless-oriented QUBO modeling guidelines for QA-assisted resource allocation. We first present a beginner-friendly toy example in which a terrestrial-aerial resource-allocation problem with a fractional signal-to-interference-plus-noise ratio (SINR) maximization objective is reformulated into a two-bit QUBO via Boolean interpolation, converted into an Ising representation, and decoded back into a wireless scheduling decision. The same scheduling is applicable to UAV, aircraft, satellite, and terrestrial links in SAGIN systems. We then review the QA foundations and solver ecosystems, including binary quadratic, discrete quadratic, constrained quadratic, and hybrid models from the perspective of NG wireless optimization. Furthermore, we discuss both binary and discrete encoding, penalty-based constraint conversion, embedding overhead, solution decoding, and feasibility verification. A QUBO Compatibility Index (QCI) is introduced as a lightweight diagnostic metric for assessing the associated QUBO formulation readiness, rather than a predictor of runtime or solution quality. Finally, a case study on max-min uplink power control in cell-free massive MIMO illustrates how a fairness-oriented wireless problem can be discretized, constrained, converted into the concept of a QUBO, and evaluated against classical baselines. A host of key challenges and future directions are also discussed.
\end{abstract}

\maketitle

\section{Introduction}
\IEEEPARstart{N}{ext-Generation} (NG) Networks are expected to support massive connectivity, high throughput, ultra-low latency, and energy-efficient operation \cite{Hanzo_road_6G, 2020_OJCOMS_Chwdhury_6G_survey, ammar2021user}. 
These requirements are particularly demanding in 
heterogeneous wireless systems such as uncrewed aerial vehicle (UAV)-assisted networks, vehicular networks, and space-air-ground integrated networks (SAGINs),
where high mobility, rapidly varying channels, intermittent connectivity, and delay-sensitive services require fast and reliable resource-allocation decisions \cite{2024_OJCOMS_SAGIN, 2026_OJCOMS_SAGIN, 2018_COMST_Liu_SAGIN_survey, 2022_IWC_Liu_SAGIN}.
Meeting these requirements necessitates the dynamic allocation of spectrum, transmit power, time-frequency resources, and beamforming degrees of freedom under stringent latency and energy budgets, while supporting heterogeneous services and network slices. These resource-allocation tasks are commonly formulated as constrained optimization problems that mix continuous control variables (e.g., transmit power, beamformers) with discrete on-off scheduling and assignment decisions (e.g., user-RB association, antenna selection)~\cite{mai2018pilot, liu2016user, 10756787}. In realistic deployments, the resulting problems are frequently of nonconvex and large-scale nature~\cite{2023_TMC_JointOpt, 2025_OJVT_Berra_Largescale}, motivating a range of approximate solvers, including convex relaxations, decomposition methods, heuristics, and metaheuristic optimization ~\cite{2025_Heuristic_Meta_Compare, van2024joint, 2026_OJCOMS_Awad_metasheutistic}. However, the increasing dimensionality and combinatorial structure of some wireless resource-allocation instances continue to raise practical concerns regarding time-to-solution, constraint satisfaction, and robustness under dynamic network conditions~\cite{Afifi_2025_RL}. While deep learning-based methods enable prompt inference once trained, their performance is highly dependent on the availability of large datasets and tends to degrade when network conditions deviate from the training distribution~\cite{2024_Mag_DL_limit,11177505, 2024_OJCOMS_Celik_ML, 2024_OJCOMS_Aouedi_FL}.

Recently, an alternative computational paradigm has been explored for tackling NP-hard combinatorial resource-optimization problems in wireless communications. The central idea is to encode the target problem into a QUBO formulation or, equivalently, an Ising model and then obtain approximate solutions using Ising machines. A representative realization of a quantum Ising machine is the quantum annealer, which implements QA motivated by adiabatic quantum evolution, provided that the problem can be expressed in Ising form~\cite{albash2018adiabatic}. In this setting, QA acts as a hardware-based metaheuristic that seeks low-energy states of the corresponding Ising model, thereby providing native support for binary quadratic optimization. Because many network-control tasks contain discrete components or can be discretized, QA has been explored as an alternative computational primitive for specific combinatorial subproblems in wireless communications and networking~\cite{Krikidis2024ConfigSelection,Zheng2025HOBO}. Leveraging quantum tunneling and parallel state evolution \cite{2025_PhysRev_QA_pros}, QA can efficiently traverse potential energy barriers and mitigate the risk of convergence to local minima \cite{qa_for_optimize}. Moreover, QA has been realized on commercial quantum hardware, such as D-Wave systems, underscoring its practicality and accessibility for real-world deployment~\cite{okada2019improving}. However, QA cannot be viewed as a universal replacement for classical methods: its performance depends strongly on the problem structure, encoding choices, constraint handling, and the limitations imposed by connectivity, noise, and precision of current devices. Consequently, a key practical question for wireless researchers and engineers is how to determine when QA is an appropriate tool, how to translate wireless optimization models into QA-compatible formulations, and how to evaluate results fairly against classical baselines.

Despite growing interest \cite{Ito2020CIMResource, Obata2024SBMResource}, the application of QA and Ising-machine-based optimization of NG-network resource allocation remains fragmented. Most existing contributions focus on specific wireless optimization tasks, such as resource allocation using coherent Ising machines, NOMA resource allocation \cite{Otsuka2024CIMNOMA}, 5G resource allocation using simulated bifurcation, reconfigurable MIMO antenna configuration selection \cite{Krikidis2024ConfigSelection}, or constrained higher-order binary optimization in wireless systems \cite{Zheng2025HOBO}. These contributions are important demonstrations of Ising-machine-based optimization in wireless communications. However, they do not provide unified wireless-oriented guidelines for explaining when QA/QUBO modeling is appropriate, how wireless variables and constraints should be encoded, how binary quadratic model (BQM), discrete quadratic model (DQM), and constrained quadratic model (CQM) solver interfaces should be selected, and how hardware-aware metrics should be reported for reproducible benchmarking. To make this modeling gap concrete, this paper first introduces a minimal resource-allocation example in a terrestrial-aerial communication network, where two binary transmit-or-silent decisions interact through interference-coupled fractional signal-to-interference-plus-noise-ratio (SINR) terms. This toy example traces the complete conversion from a wireless SINR objective to a two-bit QUBO, its equivalent Ising form, and the decoded scheduling action, thereby providing the intuition needed before moving to our general modeling guidelines and to our the larger-scale cell-free massive MIMO case study. Building on this tutorial example, this paper addresses the broader knowledge gap by developing a QUBO modeling guide for QA-assisted wireless resource allocation. Again, our objective is not to present QA as a universal replacement for classical optimization, but to clarify the modeling steps, limitations, and reporting requirements that determine whether a wireless optimization problem may be meaningfully transformed into a QA-compatible binary quadratic model. The novelty of this paper is boldly contrasted against representative related studies in Table~\ref{table:novelty_contrast}. Motivated by these gaps, the main contributions of this paper are summarized as follows:

\begin{table*}[t]
\renewcommand{\arraystretch}{2}
\centering
\caption{Contrasting the proposed QA/QUBO guideline against representative Ising-machine and QA-based wireless studies.}
\label{table:novelty_contrast}
\resizebox{\textwidth}{!}{
\begin{tabular}
{|l|c|c|c|c|c|c|}

\hline
\textbf{Feature}
& Ito, H. \textit{et al.} \cite{Ito2020CIMResource}
& Otsuka, T. \textit{et al.} \cite{Otsuka2024CIMNOMA}
& Obata, H. \textit{et al.} \cite{Obata2024SBMResource}
& Krikidis, I. \textit{et al.} \cite{Krikidis2024ConfigSelection}
& Zheng, G. \textit{et al.} \cite{Zheng2025HOBO}
& \makecell[c]{\textbf{This} \\ \textbf{study}} \\
\hline\hline

Wireless resource allocation
& \checkmark & \checkmark & \checkmark & \checkmark & \checkmark & \checkmark \\ \hline

Ising/QA-based formulation
& \checkmark & \checkmark & \checkmark & \checkmark & \checkmark & \checkmark \\ \hline

Problem-specific optimization algorithm
& \checkmark & \checkmark & \checkmark & \checkmark & \checkmark & \checkmark \\ \hline

Step-by-step toy example
& -- & -- & -- & -- & -- & \checkmark \\ \hline

QA/QUBO modeling guideline
& -- & -- & -- & -- & $\triangle$ & \checkmark \\ \hline

Solver ecosystem
& -- & -- & -- & -- & -- & \checkmark \\ \hline

Variable encoding guideline
& -- & $\triangle$ & $\triangle$ & $\triangle$ & $\triangle$ & \checkmark \\ \hline

Constraint handling (true-QUBO)
& -- & $\triangle$ & $\triangle$ & $\triangle$ & \checkmark & \checkmark \\ \hline

QUBO-readiness (QCI)
& -- & -- & -- & -- & -- & \checkmark \\ \hline

Embedding-aware metrics
& -- & -- & -- & $\triangle$ & $\triangle$ & \checkmark \\ \hline

Cell-free massive MIMO case study
& -- & -- & -- & -- & -- & \checkmark \\ \hline

Challenges and future directions
& -- & -- & -- & $\triangle$ & $\triangle$ & \checkmark \\ \hline

\end{tabular}
}
\vspace{1mm}
\begin{flushleft}
\footnotesize
\emph{Legend:} $\checkmark$ denotes explicit coverage, $\triangle$ denotes partial or problem-specific coverage, and -- denotes that the feature is not a main focus of the corresponding work.
\vspace{-0.5cm}
\end{flushleft}
\end{table*}



\begin{itemize}
    \item The paper develops structured guidelines for translating wireless resource-allocation problems into QA-compatible optimization models. The guidelines cover the associated problem classification, solver-interface selection, binary and discrete encoding, penalty-based constraint handling, embedding considerations, solution decoding, and feasibility verification. In contrast to generic QA tutorials, the discussion is tailored to wireless problems characterized by coupled interference, mixed discrete-continuous decisions, fairness constraints, and latency-sensitive operation.

    \item To provide an end-to-end tutorial bridge between wireless resource allocation and the QA workflow, the paper presents a rudimentary resource-allocation example, which may be applied to more sophisticated UAV-assisted, aircraft-assisted, and satellite-assisted SAGIN links. Starting from an interference-coupled fractional SINR objective, the example derives an exact QUBO formulation, maps it to the corresponding Ising representation, and decodes the lowest-energy state into the final scheduling decision.

    \item {\color{black}To assess how naturally a wireless optimization problem fits a binary quadratic representation, we introduce the QUBO Compatibility Index (QCI) as a lightweight formulation-level diagnostic. Its practical use is demonstrated through before/after reformulation comparisons in both the toy example and the cell-free massive MIMO case study, illustrating how QCI identifies the extent to which a formulation has been transformed into a QUBO-native form. We deliberately position QCI as a screening metric rather than a predictor of runtime, solution quality, or scalability. Accordingly, its interpretation is complemented by post-encoding and hardware-aware quantities, including the logical-variable and auxiliary-variable counts, QUBO graph density, coefficient scaling, physical-qubit usage, chain statistics, feasibility rate, and post-processing requirements.}
    
    \item To exemplify the employment of the proposed guidelines, we provide a specific case study of the max-min uplink power-control in cell-free massive MIMO. This illustrates how a wireless fairness problem can be discretized, encoded, constrained, and converted into a binary quadratic model suitable for QA-based solution. Furthermore, we identify key challenges and future research directions for QA-assisted wireless optimization, including hardware scalability, noise sensitivity, QUBO representation fidelity, dynamic network adaptation, and standardized benchmarking.
\end{itemize}
The rest of this paper is organized as follows: Section~II introduces a beginner-friendly terrestrial-aerial communication toy example that traces the modeling path from a wireless SINR objective to a QUBO matrix, an Ising representation, and a decoded scheduling decision. Section~III reviews the theoretical foundations of QA and provides an overview of the available solver ecosystems. Section~IV addresses practical considerations related to QUBO embedding and variable encoding in wireless optimization problems. Section~V presents a detailed case study based on the application of QA to a max-min fairness power-control problem in cell-free massive MIMO systems. Section~VI discusses the key challenges and outlines future research directions for the integration of QA into next-generation networks. Finally, Section~VII summarizes the main insights and concludes the paper.

\section{Quantum Annealing: A Beginner-Friendly QUBO Example}
\label{sec:toy_example}

\subsection{A Terrestrial-Aerial Communication Toy Example: From SINR Maximization to QUBO}

\begin{figure*}[t]
    \centering
    \includegraphics[width=\linewidth]{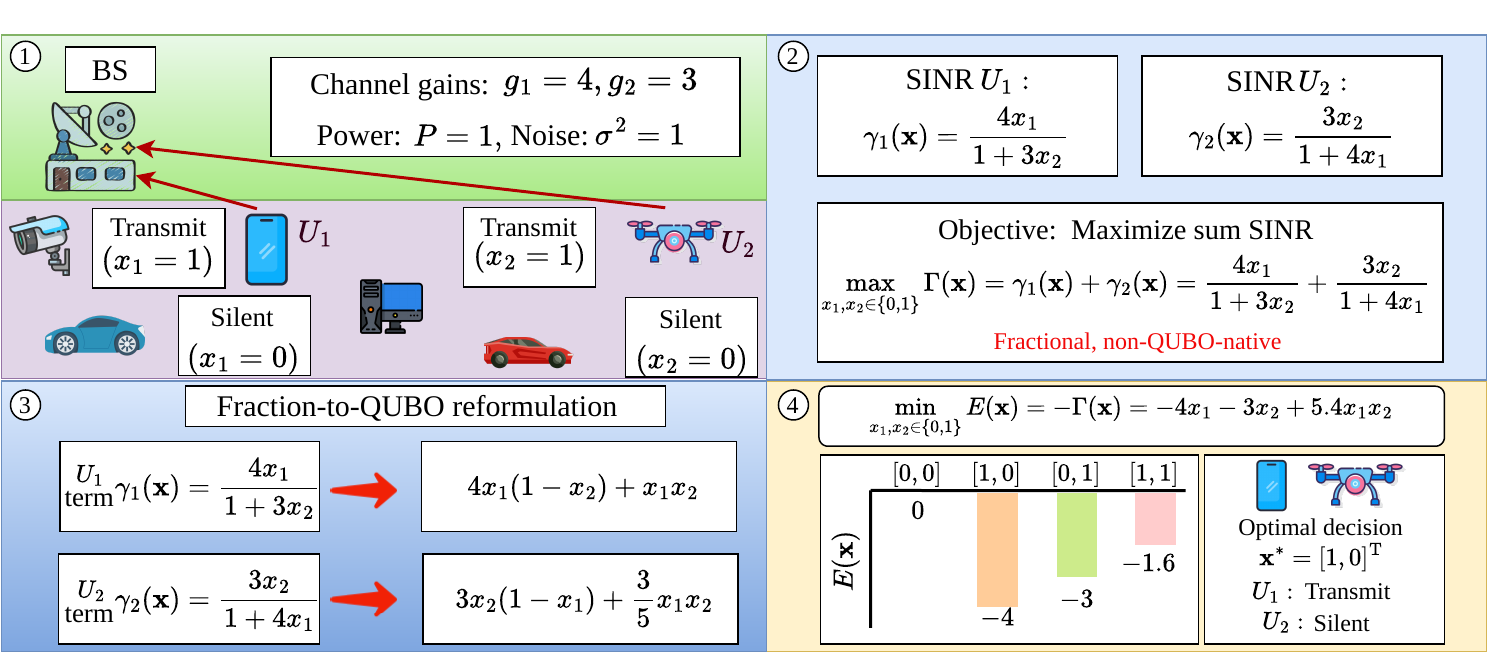}
    \caption{A toy example of an exact SINR-to-QUBO reformulation for scheduling in terrestrial-aerial communication networks. 
    (1) A ground user equipment, denoted by $U_1$, and an aerial user equipment, denoted by $U_2$, compete for a single resource block and cause mutual interference at the BS. 
    (2) The original objective is a fractional, non-QUBO-native sum-SINR maximization problem. 
    (3) Since the scheduling variables are binary, each fractional SINR term can be exactly rewritten as a quadratic binary expression via Boolean interpolation. 
    (4) The resulting QUBO energy landscape shows that the minimum-energy state is $\mathbf{x}^{\star}=[1,0]^T$, corresponding to scheduling the stronger $U_1$ while keeping $U_2$ silent for this toy instance.}
    \label{fig:toy_example}
    \vspace{-0.2cm}
\end{figure*}

Before presenting the general QA workflow, we first introduce a rudimentary resource-allocation example to illustrate how a wireless optimization problem can be transformed into a QUBO form. 
Fig.~\ref{fig:toy_example} illustrates the example in a terrestrial-aerial communication network, where a base station (BS) serves ground user equipment $U_1$ and aerial user equipment $U_2$ over a single time-frequency resource block (RB). However, the mathematical model should be interpreted more generally as a two-node scheduling abstraction. Specifically, the receiver may represent a roadside unit (RSU), a UAV relay, an aircraft gateway, or a satellite, ground-station gateway. Likewise, the two transmitters may represent terrestrial V2X vehicles, UAVs in an aerial access network, aircraft in an aeronautical network, or satellite/terrestrial nodes in a SAGIN. This interpretation is aligned with the practical multi-layer SAGIN architecture, where terrestrial, aerial, and satellite segments are integrated under heterogeneous and time-varying resource-management constraints \cite{2018_COMST_Liu_SAGIN_survey, 2022_IWC_Liu_SAGIN}.
In this example, each user makes a binary transmission decision:
\[
x_k =
\begin{cases}
1, & \text{if user } U_k \text{ transmits on the RB considered},\\
0, & \text{otherwise}.
\end{cases}
\]
For simplicity, we assume normalized transmit power of $P=1$, normalized noise power of $\sigma^2=1$, and channel gains $g_1=4$ and $g_2=3$. When both users transmit simultaneously over the same RB, they interfere with each other at the BS. Hence, the SINR values of $U_1$ and $U_2$ can be written as
\begin{equation}
    \begin{aligned}
        & \gamma_1(\mathbf{x}) = \frac{g_1x_1}{\sigma^2+g_2x_2} = \frac{4x_1}{1+3x_2}, \\
        & \gamma_2(\mathbf{x}) = \frac{g_2x_2}{\sigma^2+g_1x_1} = \frac{3x_2}{1+4x_1},
    \end{aligned}
\end{equation}
where $\mathbf{x}=[x_1,x_2]^T$. In this toy example, the resource-allocation objective is to maximize the sum-SINR:
\begin{equation}
    \max_{x_1,x_2\in\{0,1\}}
    \Gamma(\mathbf{x})
    =
    \gamma_1(\mathbf{x})+\gamma_2(\mathbf{x})
    =
    \frac{4x_1}{1+3x_2}
    +
    \frac{3x_2}{1+4x_1}.
    \label{2}
\end{equation}
Although this objective is fractional and it is not directly in QUBO form, its binary structure makes an exact quadratic reformulation possible. Specifically, for binary variables $x,y\in\{0,1\}$ and constants $a,b>0$, we have
\begin{equation}
\label{equa_3}
    \frac{a x}{1+b y}
    =
    a x(1-y)+\frac{a}{1+b}xy.
\end{equation}
This identity is exact because $y$ can only take two values. If $y=0$, the right-hand side becomes $ax$; if $y=1$, it becomes $ax/(1+b)$. Applying this identity to the above two SINR terms gives
\begin{equation}
    \begin{aligned}
        & \frac{4x_1}{1+3x_2}=4x_1(1-x_2)+x_1x_2,\\
        & \frac{3x_2}{1+4x_1}=3x_2(1-x_1)+\frac{3}{5}x_1x_2.
    \end{aligned}
\end{equation}
Consequently, the sum-SINR objective can be rewritten as
\begin{equation}
    \begin{aligned}
        \Gamma(\mathbf{x})
        &=
        4x_1(1-x_2)+x_1x_2
        +
        3x_2(1-x_1)+\frac{3}{5}x_1x_2 \\
        &=
        4x_1+3x_2-5.4x_1x_2 .
    \end{aligned}
\end{equation}
Since QUBO is typically formulated as a minimization problem, maximizing the sum-SINR $\Gamma(\mathbf{x})$ is equivalent to minimizing its negated version:
\begin{equation}
    \min_{x_1,x_2\in\{0,1\}}
    E(\mathbf{x})
    =
    -\Gamma(\mathbf{x})
    =
    -4x_1-3x_2+5.4x_1x_2 .
\end{equation}
In physically plausible terms, this negation transforms the sum-SINR peak into a trough. This is already in QUBO form. In scalar form, the QUBO coefficients are
$
q_1=-4, q_2=-3, q_{12}=5.4,
$ hence, we have
$
E(\mathbf{x})=q_1x_1+q_2x_2+q_{12}x_1x_2 .
$
Equivalently, using the symmetric matrix convention of $E(\mathbf{x})=\mathbf{x}^{T}\mathbf{Q}\mathbf{x}$, a valid QUBO matrix representation is:
\[
\mathbf{Q}
=
\begin{bmatrix}
-4 & 2.7\\
2.7 & -3
\end{bmatrix}.
\]
The diagonal entries represent the individual benefit of scheduling each user, while the off-diagonal term represents the pairwise interference penalty. The negative first-order coefficients encourage transmission, whereas the positive second-order coefficient penalizes simultaneous transmission. All possible configurations can be explicitly enumerated as:
\[
\renewcommand{\arraystretch}{1.5}
\begin{array}{c|c|c|c|c}
x_1 & x_2 & \text{Calculation of } \Gamma(\mathbf{x}) & \Gamma(\mathbf{x}) & E(\mathbf{x})=-\Gamma(\mathbf{x})\\
\hline
0 & 0 & 4(0)+3(0)-5.4(0)(0) & 0   & 0\\
1 & 0 & 4(1)+3(0)-5.4(1)(0) & 4   & -4\\
0 & 1 & 4(0)+3(1)-5.4(0)(1) & 3   & -3\\
1 & 1 & 4(1)+3(1)-5.4(1)(1) & 1.6 & -1.6
\end{array}
\]
The minimum energy is obtained at
$
\mathbf{x}^{\star}=[1,0]^T.
$
Therefore, the optimal decision is to schedule $U_1$ on the RB considered, while keeping $U_2$ silent. This result is intuitive: $U_1$ has the stronger channel, and simultaneous transmission by both users causes sufficient mutual interference to reduce the sum-SINR. This illustrates the core principle of wireless QUBO modeling: binary variables represent allocation decisions, linear terms capture individual gains, and quadratic terms capture interactions such as interference, or resource conflicts.

\subsection{From the Toy QUBO to a Decoded Scheduling Decision}

We now illustrate how the QUBO energy is interpreted by a QA workflow. The energy function is first viewed as a logical graph, then expressed in Ising spin variables~\cite{shirai2023spin}, and finally decoded back into the original wireless scheduling decision. The QUBO graph associated with $E(\mathbf{x})$ has two logical nodes and one logical edge: $\mathcal{V}_Q=\{x_1,x_2\}$, $\mathcal{E}_Q=\{(x_1,x_2)\}$. The node biases are $-4$ and $-3$, while the edge weight is $5.4$. Thus, the two nodes represent the scheduling decisions of $U_1$ and $U_2$, while the edge represents the interference coupling between them. To express the problem in Ising form~\cite{bybee2023efficient}, the binary variables are mapped to spin variables $s_i\in\{-1,+1\}$ 
\begin{equation}
    s_i=2x_i-1, \qquad x_i=\frac{s_i+1}{2}.
\end{equation}
Substituting this mapping into the QUBO energy gives
\begin{equation}
    \begin{aligned}
    E(\mathbf{s})
    &=
    -4\frac{s_1+1}{2}
    -3\frac{s_2+1}{2}
    +5.4\frac{(s_1+1)(s_2+1)}{4} \\
    &=
    1.35s_1s_2-0.65s_1-0.15s_2-2.15 .
    \end{aligned}
\end{equation}
The constant offset $-2.15$ does not affect the minimizer. Hence, the corresponding Ising parameters are
$
h_1=-0.65, h_2=-0.15, J_{12}=1.35.
$
For this two-variable example, the hardware embedding is trivial, whenever two directly connected physical qubits are available. In that case, one may simply map $s_1\mapsto q_1$ and $s_2\mapsto q_2$, where $q_1$ and $q_2$ are connected physical qubits. Hence, no qubit chain is required in this toy example. Larger wireless QUBOs generally require minor embedding~\cite{minor_embedding}, as discussed in the next section. The end-to-end bit-level trace is summarized in Table~\ref{tab:toy_bit_flow}. The table follows the same sequence as that used in the QA workflow: the wireless decision is represented by binary variables, where the binary objective is encoded by a QUBO matrix and graph. To elaborate, the QUBO is converted into Ising parameters, and the lowest-energy spin assignment is decoded back into the original scheduling decision.
\begin{table}[t]
\centering
\caption{End-to-end bit-level trace of the toy QUBO.}
\label{tab:toy_bit_flow}
\renewcommand{\arraystretch}{1.8}
\begin{tabular}{m{0.2\linewidth}|m{0.7\linewidth}}
\hline
\textbf{Stage} & \textbf{Toy-instance realization} \\
\hline
Wireless bits
&
$\mathbf{x}=[x_1,x_2]^T$, where $x_k=1$ means that $U_k$ transmits on the RB considered. \\
\hline
QUBO matrix
&
$E(\mathbf{x})=\mathbf{x}^{T}\mathbf{Q}\mathbf{x}$, with
$\mathbf{Q}=
\begin{bmatrix}
-4 & 2.7\\
2.7 & -3
\end{bmatrix}$.
\\
\hline
QUBO graph
&
Two logical nodes having biases $-4$ and $-3$, connected by a logical edge with weight $5.4$. The edge captures the interference penalty. \\
\hline
\makecell[l]{Ising\\parameters}
&
Using $s_i=2x_i-1$, the equivalent Ising form is
$E(\mathbf{s})=1.35s_1s_2-0.65s_1-0.15s_2-2.15$.
Thus, $h_1=-0.65$, $h_2=-0.15$, and $J_{12}=1.35$. \\
\hline
\makecell[l]{Best spin\\assignment}
&
The lowest-energy spin assignment is
$\mathbf{s}^{\star}=[+1,-1]^T$. \\
\hline
\makecell[l]{Decoded\\wireless action}
&
Using $x_i=(s_i+1)/2$, the spin assignment is decoded as
$\mathbf{x}^{\star}=[1,0]^T$.
Hence, $U_1$ transmits, while $U_2$ remains silent. \\
\hline
\end{tabular}
\end{table}
For the last two rows of Table~\ref{tab:toy_bit_flow}, the spin assignment $\mathbf{s}^{\star}=[+1,-1]^T$ is decoded as $\mathbf{x}^{\star}
= \left[\frac{1+1}{2}, \frac{-1+1}{2} \right]^T = [1,0]^T$. This recovered binary vector determines the wireless action of $U_1$: transmit and $U_2$: silent. This toy example illustrates the basic QA/QUBO modeling workflow: binary variables encode wireless decisions, linear QUBO terms capture scheduling gains, and quadratic couplings represent interference. {\color{black}
The toy example also illustrates a useful formulation-level distinction.
Before Boolean interpolation, the objective in~(\ref{2}) contains two
fractional SINR terms, neither of which is directly expressible as a binary
linear or pairwise-quadratic monomial. After applying identity~(\ref{equa_3}), the
reformulated objective $-4x_1-3x_2+5.4x_1x_2$ contains only QUBO-native terms. Thus, the fraction of QUBO-native terms changes from $0/2$ in the original representation to $3/3$ after reformulation. It should be emphasized that this ratio measures formulation compatibility rather than problem size or computational difficulty. In larger wireless problems, increasing the number of users, power levels, or auxiliary variables may substantially enlarge the resulting QUBO while the post-reformulation fraction of QUBO-native terms can still remain equal to one. This simple before/after comparison motivates the formulation-level compatibility measure introduced formally as the QUBO Compatibility Index (QCI) in Section~\ref{sec: solver_ecos}.} The example is deliberately designed to be small so that every bit, coefficient, spin assignment, and decoded scheduling action can be inspected directly. We next generalize this procedure by reviewing the QA workflow, the Ising/QUBO interface, and the solver models used for larger wireless resource-allocation problems. 


\section{Background and Solver Ecosystems of QA}
\begin{figure*}[t]
    \centering
    \includegraphics[width=\linewidth]{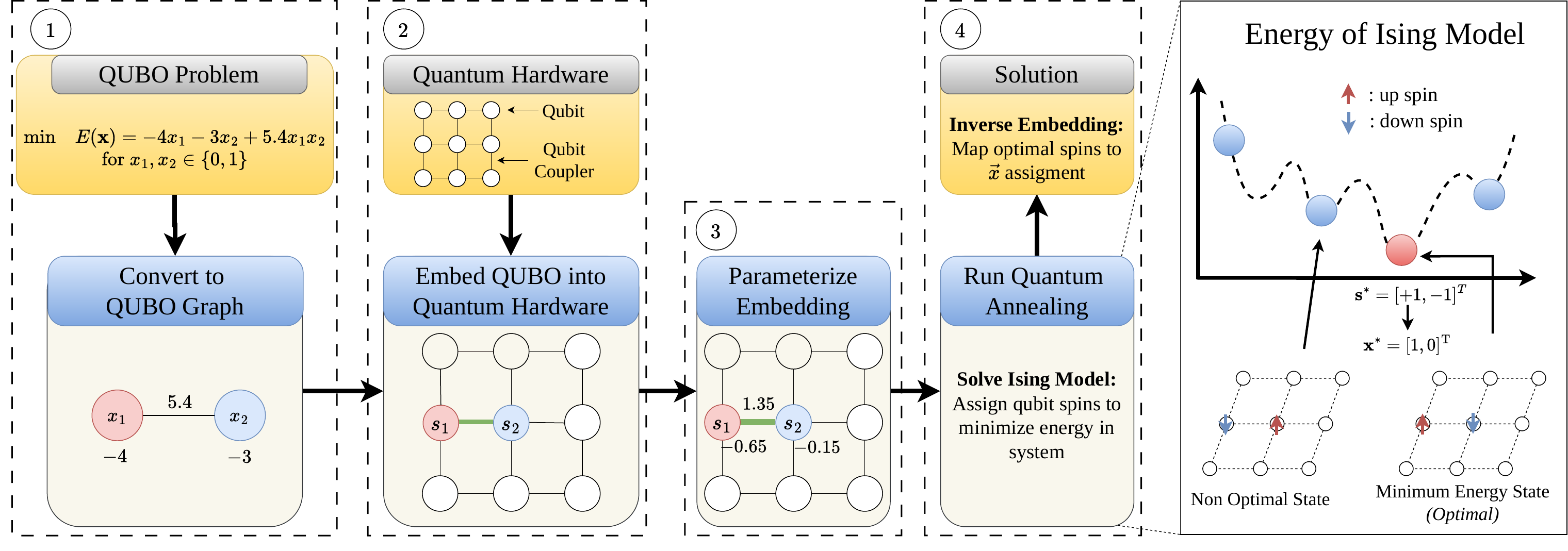}
    \caption{End-to-end workflow of QA for solving a wireless QUBO, shown as the general counterpart of the two-user toy trace in Section~\ref{sec:toy_example}. In the toy instance, the QUBO block corresponds to $E(\mathbf{x})=-4x_1-3x_2+5.4x_1x_2$, the QUBO graph consists of two logical nodes and one interference edge, the parameterization step gives the Ising coefficients $h_1=-0.65$, $h_2=-0.15$, and $J_{12}=1.35$, and the lowest-energy spin assignment is decoded as the scheduling decision $\mathbf{x}^{\star}=[1,0]^T$. For larger wireless resource-allocation problems, the same workflow applies with larger QUBO graphs, nontrivial embedding, and additional feasibility post-processing.}
    \vspace{-0.3cm}
    \label{fig:qa_workflow}
\end{figure*}

\subsection{Foundations of Quantum Annealing}
QA generalizes simulated annealing to quantum dynamics, providing a physics-inspired heuristic for high-dimensional combinatorial optimization. It is motivated by the adiabatic theorem, under which a system initialized in the ground state of a simple Hamiltonian remains in the instantaneous ground state during sufficiently slow evolution toward the problem Hamiltonian. Formally, QA implements a time-dependent Hamiltonian\cite{albash2018adiabatic} of the form
\begin{equation}
H(t) = A(t)H_0 + B(t)H_p, \quad t \in [0,\tau],
\end{equation}
where $H_0$ is a driver Hamiltonian that induces quantum fluctuations and prepares an easily accessible initial state, and $H_p$ encodes the optimization objective (and any constraints via penalty terms). The annealing schedules $\{A(t),B(t)\}$ are typically chosen to be monotone with boundary conditions $A(0)=1$, $B(0)=0$ and $A(\tau)=0$, $B(\tau)=1$. Ideally, if the annealing time $\tau$ exceeds a critical threshold dictated by the minimum spectral gap $\Delta_{\min}$, the system follows the ground state trajectory and {converges to the global optimum}. In practice, noise, decoherence, and finite hardware runtimes prevent perfect adiabaticity. Modern QA systems therefore operate heuristically: the quantum processor is repeatedly annealed and measured to sample low-energy solutions, often followed by classical post-processing, such as local search or constraint repair to improve solution quality. The mathematical model for QA hardware is the {Ising Hamiltonian}\cite{lucas2014ising}:
\begin{equation}
H_{\text{Ising}}(\{\sigma_i\}) = \sum\nolimits_i h_i \sigma_i^z + \sum\nolimits_{i<j} J_{ij} \sigma_i^z \sigma_j^z,
\end{equation}
where $\sigma_i^z$ represents Pauli matrices operating on the qubit $i$, $h_i$ are local biases, and $J_{ij}$ are pairwise couplings. Accordingly, the classical version of the Ising model can be written as 
 \begin{equation}
     H_{\mathrm{ising}}(\{s_i\}) = \sum\nolimits_i h_is_i + \sum\nolimits_{i< j}J_{ij} s_i s_j,
 \end{equation}
where $s_i \in \{-1,+1\}$ denotes the $i$-th spin of a qubit. 
To facilitate integration with digital optimization frameworks, the Ising model is commonly transformed into a QUBO problem via the mapping $s_i = 2x_i - 1$, where $x_i \in \{0,1\}$. The resultant QUBO formulation is then given by
\begin{equation}
H_{\text{QUBO}}(x) = \sum\nolimits_i a_i x_i + \sum\nolimits_{i<j} b_{ij} x_i x_j,
\end{equation}
which is algebraically equivalent to the Ising Hamiltonians, but uses binary variables that are more amenable to classical preprocessing and hybrid workflows. Because QA hardware natively relies on quadratic binary forms, QUBO serves as the interface between high-level optimization problems and quantum annealers. The Ising-QUBO equivalence provides a universal bridge, enabling wireless optimization tasks to be encoded in a form compatible with quantum solvers. 

Figure~\ref{fig:qa_workflow} generalizes the bit-level trace of the toy example in Section~\ref{sec:toy_example}. For the two-user instance, the ``QUBO problem'' block corresponds to  $E(\mathbf{x})=-4x_1-3x_2+5.4x_1x_2$, and the ``QUBO graph'' block contains two logical nodes, $x_1$ and $x_2$, connected by one interference edge of weight $5.4$. The parameterization step corresponds to the binary-to-spin mapping $s_i=2x_i-1$, which gives the Ising coefficients $h_1=-0.65$, $h_2=-0.15$, and $J_{12}=1.35$. For this toy instance, embedding is trivial when two connected physical qubits are available, and the lowest-energy spin assignment $\mathbf{s}^{\star}=[+1,-1]^T$ is decoded back as $\mathbf{x}^{\star}=[1,0]^T$. Thus, Fig.~\ref{fig:qa_workflow} should be read as the general version of the concrete toy pipeline: QUBO formulation, graph construction, hardware embedding, parameterization, annealing, and inverse decoding. For larger wireless resource-allocation problems, the same workflow applies, but the QUBO graph is typically much larger and may be dense. The embedding step may then require chains of physical qubits and classical post-processing to repair infeasible or broken-chain solutions. Hence, the toy example provides the smallest inspectable instance of the workflow, while Fig.~\ref{fig:qa_workflow} summarizes the corresponding general procedure.

\begin{remark}
The transition from wireless optimization problems to QA hardware relies critically on the {binary-quadratic reformulation} of decision variables. Through the Ising/QUBO representation, resource-allocation decisions are encoded as quadratic energy functions that can be processed by quantum annealers or hybrid solvers.
\end{remark}

\subsection{Solver Ecosystems}
\label{sec: solver_ecos}
D-Wave's Ocean and Leap ecosystem~\cite{Osaba2025NPHS} provides several modeling interfaces for optimization problems having diverse structural properties. It is important to distinguish between models that are native to direct quantum processing unit (QPU) sampling and higher-level models that are typically handled by quantum-classical hybrid solvers. According to D-Wave documentation \cite{DWaveModels}, direct QPU submissions are based on unconstrained binary quadratic models, such as QUBO and Ising models. By contrast, DQM and CQM provide higher-level modeling abstractions and are typically solved through hybrid quantum-classical solvers \cite{DWaveEmbedding}.

    $\bullet$ \textbf{Binary Quadratic Model (BQM):} A BQM is an unconstrained quadratic model over binary or spin variables and encodes the Ising/QUBO formulations used by D-Wave samplers \cite{DWaveModels}. It is therefore suitable for wireless problems that can be expressed using binary decision variables and pairwise interactions, such as user association, link activation, antenna selection, and binary scheduling. Since BQM/QUBO models are unconstrained, feasibility constraints must be incorporated through penalty term~\cite{DWaveModels}.

    $\bullet$ \textbf{Discrete Quadratic Model (DQM):} A DQM models optimization problems with finite-valued discrete variables using objective terms of degree at most two~\cite{DWaveModels}. Since each discrete variable can be converted to binary form through one-hot encoding, DQM is useful for wireless tasks such as channel, modulation, beam-index, and discrete power selection. However, DQM is a high-level hybrid-solver interface rather than a direct hardware-level QUBO embedding.

    $\bullet$ \textbf{Constrained Quadratic Model (CQM):} A CQM supports objectives and constraints involving binary, integer, and real variables. Furthermore, it allows constraints to be modeled explicitly \cite{DWaveModels}. This is convenient for wireless problems having power-budget, scheduling, or fairness constraints. Nevertheless, CQM is also a hybrid-solver modeling interface, rather than a direct QPU formulation. Moreover, real-valued variables are currently not supported in quadratic interactions \cite{DWaveModels}, so continuous wireless variables may still require discretization, linearization, or problem-specific reformulation.
    
To provide a preliminary indication of whether a wireless optimization problem is naturally close to a QUBO representation, we introduce the QUBO Compatibility Index (QCI). Consider a wireless optimization problem written in a fixed canonical algebraic representation before QUBO conversion. Let $\mathcal{T}$ denote the set of elementary modeling terms appearing in the objective and constraint functions after polynomial expansion, excluding constant offsets. A term is referred to as QUBO-native if it can be represented as a binary linear or binary quadratic term, such as
\begin{equation}
    a_i x_i
    \quad \textrm{or} \quad
    a_{ij}x_i x_j,
    \qquad x_i,x_j\in\{0,1\},
\end{equation}
without requiring continuous-variable discretization, nonlinear approximation, higher-order reduction, slack-variable introduction, or auxiliary-variable introduction. This definition follows the native QUBO structure used in D-Wave documentation~\cite{DWaveModels, DWaveReformulating}, where the objective contains only linear binary terms and pairwise quadratic binary interactions. Let $\mathcal{T}_{\mathrm{nat}}\subseteq\mathcal{T}$ be the set of such QUBO-native terms. We define
\begin{equation}
    \mathrm{QCI}
    =
    \frac{
    \vert\mathcal{T}_{\mathrm{nat}}\vert
    }{
    \vert\mathcal{T}\vert
    },
    \label{eq:qci}
\end{equation}
where $\vert\cdot\vert$ represents cardinality. The before/after term counting illustrated by the toy example in
Section~II is a simple instance of this definition: QCI can be evaluated
both before and after reformulation to indicate how much of the original
problem structure is directly compatible with QUBO modeling. A high QCI indicates that a larger fraction of the problem structure is already compatible with binary quadratic modeling. Conversely, a low QCI suggests that substantial reformulation may be required before the problem may be deemed to be a QUBO. Such a reformulation action may include discretizing continuous variables~\cite{arai2023effectiveness}, replacing nonlinear functions by quadratic approximations~\cite{viet2026advanced}, quadratizing higher-order binary products~\cite{mandal2020compressed}, introducing slack variables for inequalities~\cite{lucas2014ising}, or adding auxiliary variables for product linearization~\cite{verma2020optimal}. D-Wave documentation also notes that higher-order models are typically reduced to quadratic form for sampling \cite{DWaveModels}. However, QCI should not be interpreted as a standalone measure of QA performance. Because the QPU topology is not fully connected, the logical QUBO graph must be minor-embedded \footnote{Minor-embedding is the process of mapping an arbitrary problem graph (the \emph{logical graph}) onto the fixed physical layout of a quantum processor (the \emph{hardware graph}).} into the hardware graph; this process may require representing a specific logical variable by a chain of physical qubits, and the choice of embedding can affect solution quality \cite{DWaveEmbedding}. Therefore, QCI should be reported together with post-encoding and hardware-aware quantities, such as the number of logical binary variables $N_b$, the number of auxiliary variables $N_{\mathrm{aux}}$, the number of nonzero quadratic couplings $\vert E_Q\vert $, the QUBO graph density defined as $d_Q = \frac{2\vert E_Q\vert }{N_b(N_b-1)}$, the number of physical qubits after embedding, the average chain length, and the empirical feasibility rate of the sampled solutions.

\begin{remark}
QCI is a diagnostic metric defined for assessing formulation readiness for QUBO modeling. It helps identify how substantial reformulation is required before a wireless optimization problem becomes compatible with binary quadratic optimization. It does not, by itself, guarantee low embedding overhead, reduced runtime, or high solution quality. Such claims must be supported by the final QUBO size, graph density, and empirical solver performance.
\end{remark}

\subsection{Discussion: Escaping Local Minima through Quantum Tunneling}

A key difference between QA and classical SA is how they escape local minima. In SA, escape relies on thermal activation, with probability $P_{\mathrm{SA}} \sim \exp\left(-\frac{\Delta E}{k_{B}T}\right)$ for crossing an energy barrier of height $\Delta E$ at temperature $T$~\cite{kirkpatrick1983optimization}. As $T$ decreases, this probability decays exponentially, increasing the risk of trapping in suboptimal states. In contrast, QA exploits quantum tunneling as an additional escape mechanism~\cite{kadowaki1998quantum}. The dynamics of QA are governed by the transverse-field Ising Hamiltonian~\cite{stinchcombe1973ising}
\begin{equation}
    H = -\Gamma \sum\nolimits_{j=1}^{N} \sigma_j^{x} - \sum\nolimits_{i,j=1}^{N} J_{ij} \sigma_i^{z} \sigma_j^{z},
\end{equation}
where $N$ is the number of spin variables, $\Gamma$ is the transverse-field strength that controls quantum fluctuations, $\sigma_j^{x}$ and $\sigma_j^{z}$ are Pauli matrices acting on spin $j$. In the limit of small $\Gamma$, the system often admits a pair of nearly degenerate configurations, denoted as $\ket{\uparrow}$ and $\ket{\downarrow}$, which correspond to distinct local minima of the cost landscape. The transverse-field term mixes these configurations with a tunneling matrix element $\epsilon = \bra{\uparrow} H \ket{\downarrow}$, thereby lifting the degeneracy and generating an \emph{energy gap} $\Delta = 2 \epsilon$. This avoided level crossing creates a quantum tunneling channel between local minima, enabling the system to transition toward the global optimum.

From a semiclassical perspective, tunneling can be described using the Euclidean (imaginary-time) action. The tunneling amplitude is given by $A_{\text{tunnel}} \propto \exp\!\left(-\frac{S_\tau[x]}{\hbar}\right)$~\cite{isakov2016understanding}, where $\hbar$ is the reduced Planck constant and
\begin{equation}
    S_\tau[x] = \int_{0}^{\tau} d\tau' \left[ \tfrac{1}{2} m \dot{x}^{2}(\tau') + V(x(\tau')) \right].
\end{equation}
Here, $x(\tau)$ denotes the collective coordinate along the tunneling path, $\dot{x}(\tau')$ is its derivative with respect to imaginary time $\tau'$, $m$ is an effective mass parameter, and $V(x)$ represents the potential energy landscape.

Finally, the \emph{escape rate} from a metastable local minimum scales with the energy gap as~\cite{isakov2016understanding}
\[\label{eq:no7}
    W \propto \left\{
    \resizebox{0.85\hsize}{!}{$
    \begin{matrix}
    \Delta^{2} \quad \text{incoherent tunneling, open-system QA,} \\ 
    \Delta \quad \text{projected ground-state tunneling.} 
    \end{matrix}$}
    \right.
\]

These scaling relationships demonstrate that QA can successfully escape local minima even at zero temperature. Consequently, QA offers a robust mechanism for traversing complex non-convex landscapes, thereby increasing the likelihood of reaching globally optimal solutions.


\section{Practical Considerations for QUBO Embedding and Variable Encoding}

Transforming wireless resource allocation problems into QUBO form requires binary encoding of all decision variables while considering hardware connectivity and qubit limitations.
\subsection{Encoding Discrete Variables}

Discrete variables such as power levels or fairness thresholds can be encoded in one of the following ways:

$\bullet$ \textbf{One-hot encoding:} Each discrete value is assigned a dedicated binary variable, with exactly one active bit. For $P$ possible values, $P$ binary variables are required, along with a constraint that enforces exclusivity. 

$\bullet$ \textbf{Binary encoding:} The value is represented in binary format using $\lceil \log_2 (P+1) \rceil$ bits. This reduces the number of variables required, thus lowering the QUBO size. However, it introduces higher complexity in constraint design and increases the likelihood of generating invalid or ambiguous solutions without additional constraints.



\subsection{Encoding Continuous Variables via Binary Expansion}

Continuous wireless variables, such as power-control coefficients, can be approximated using fixed-point binary encoding, where each real value is decomposed into two parts:

\begin{itemize}
    \item \textbf{Integer part}: encodes the magnitude before the decimal point. 
    \item \textbf{Fractional part}: encodes the precision after the decimal point, with each bit weighted by $2^{-i}$. 
\end{itemize}

Formally, if we allocate a total of $n = n_i + n_f$ bits, where $n_i$ and $n_f$ denote the number of integer and fractional bits, respectively, then a number can be expressed as
\begin{equation}
    x \approx \sum_{k=0}^{n_i-1} b_k \, 2^k \;+\; \sum_{k=1}^{n_f} c_{k} \, 2^{-k}, 
    \quad b_k, c_k \in \{0,1\}.
    \label{eq:fixed_point}
\end{equation}

Here, the resolution of the approximation is governed by $n_f$, while the range of representable values is determined by $n_i$. Although this scheme enables continuous values to be expressed using binary variables, it increases the dimensionality of the resulting QUBO and, consequently, the complexity of embedding.




\subsection{Embedding Overhead and Qubit Constraints}

Once a QUBO is constructed, it must be mapped onto the physical QA hardware (e.g., D-Wave), which has limited qubit connectivity (e.g., Chimera or Pegasus topologies)~\cite{bunyk2014architectural}. Each logical variable may be represented by a chain of connected 
physical qubits to ensure proper coupling and interaction between variables~\cite{pelofske20244}. This mapping results in an embedding overhead, typically quantified by the \emph{average chain length}, defined as
\begin{equation}
\text{Average chain length} = 
\frac{\text{Number of physical qubits used}}
{\text{Number of logical variables}}.
\end{equation}
A high average chain length not only increases resource consumption but also raises the likelihood of chain breaks, which can degrade solution quality. Therefore, embedding-aware QUBO design, compact encoding strategies, and constraint simplification are essential to reduce average chain length and enhance execution efficiency on quantum annealers. Consequently, efficient binary encoding and low embedding overhead are essential in order to exploit QA for large-scale wireless optimization problems.


\section{Case Study: Power Control via QA in Cell-Free Massive MIMO}
\subsection{System Model and Uplink Transmission}
\begin{figure}[t]
 \centering
\includegraphics[width=0.95\linewidth , trim={0cm 0cm 0cm 0cm}, clip]{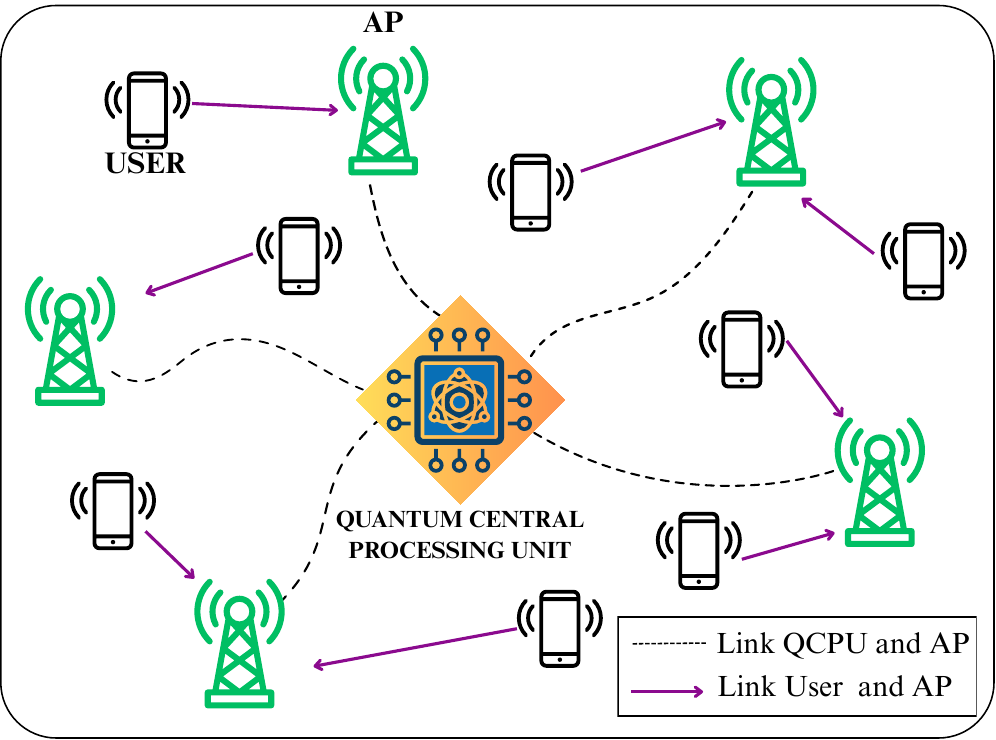}
 \caption{ Cell-free massive MIMO system with Quantum Central Processing Unit (QCPU).
}
  \label{system_model}
\end{figure}
We consider a cell-free massive MIMO network where a set of distributed access points (APs), $\mathcal{M}=\{1,\dots,M\}$, jointly serve a set of single-antenna user equipment (UE), $\mathcal{K}=\{1,\dots,K\}$, on the same time--frequency resource. Each AP is equipped with a single antenna and it is connected to a quantum central processing unit (QCPU) via backhaul/fronthaul links for resource allocation and decoding coordination \cite{2017_TWC_Hien_CfmMIMO}, as illustrated in Fig.~\ref{system_model}. Channels are assumed to be independent across AP-UE pairs and estimated during the uplink training phase. In the uplink transmission phase, the signal received at AP $m$ is
\begin{equation}\label{eq:uplink_signal}
    y_{m} = \sum\nolimits_{k=1}^{K} \sqrt{\rho_u \eta_k}\, h_{mk}\, s_k + n_{m},
\end{equation} 
where $\rho_u$ is the normalized uplink signal to noise ratio (SNR), $\eta_k\!\in\![0,1]$ is the transmit power control coefficient of UE $k$, $s_k$ is the data symbol with $\mathbb{E}[\vert s_k\vert ^2]=1$, and $n_m \sim \mathcal{CN}(0,1)$ is the receiver noise. Following the standard use-and-then-forget bounding technique \cite{2018_TGCN_Hien_CfmMIMO_EE}, the uplink SINR of UE $k$ can be expressed in closed form as in \eqref{eq:ul_sinr}.
\begin{figure*}
\vspace{-1cm}
    \begin{align}\label{eq:ul_sinr}
        \gamma^{\text{ul}}_{k} =
        \frac{\rho_u \eta_{k}\left(\sum\limits_{m=1}^M v_{mk} \right)^2 }{ 
        \rho_u \sum\limits_{\substack{k'=1 \\ 
        k'\neq k}}^K \eta_{k'} \Bigg(\sum\limits_{m=1}^M v_{mk}\frac{\beta_{mk'}}{\beta_{mk}} \vert \boldsymbol{\phi}_{k'}^{\!H}\boldsymbol{\phi}_{k}\vert \Bigg)^2
        + \rho_u \sum\limits_{k'=1}^K \eta_{k'} \sum\limits_{m=1}^M v_{mk}\beta_{mk'} 
        + \sum\limits_{m=1}^M v_{mk} },
    \end{align}
    \noindent\rule{\textwidth}{0.5pt}
\end{figure*}
Therein, $v_{mk} = \frac{\tau_p \rho_p \beta_{mk}^2}{\tau_p \rho_p \sum\nolimits_{k'=1}^K \beta_{mk'} \vert \boldsymbol{\phi}_k^{\!H}\boldsymbol{\phi}_{k'}\vert ^2 + 1}$
represents the channel estimation gain factor for AP $m$ and UE $k$. Here, $\tau_p$, $\rho_p$, and $\boldsymbol{\phi}_k$ are the uplink training duration, the normalized signal-to-noise ratio (SNR) of each pilot symbol, and the pilot sequence used by the $k$-th UE, respectively. Finally, the ergodic uplink spectral efficiency of UE $k$ (bit/s/Hz) is
\begin{equation}\label{eq:R_ul_sinr}
    R^{\text{ul}}_{k} = \log_2 \big(1+\gamma^{\text{ul}}_{k}\big).
\end{equation}
The max-min fairness ergodic rate is expressed as
\begin{subequations}\label{eq opt 1}
    \begin{align}
        && \underset{\substack{ \{\eta_{k}\} }}{\textrm{maximize}} & \quad \underset{k\in\mathcal{K}}{\min} \quad R^\text{ul}_{k} \\
        &&\textrm{subject to}&\quad 0 \leq \eta_{k} \leq 1, \forall  k\in\mathcal{K}, \label{eq opt 1 - constraint1}
    \end{align}
\end{subequations}
Although the max-min fairness appears simple, it is fundamentally nonconvex due to the fractional nature of the SINR in $R^\text{ul}_k$. The use of a minimum operator in the objective further introduces non-smoothness, making direct optimization intractable. To address these challenges, we reformulate the problem into epigraph form, this leads to the following equivalent formulation:
\begin{equation}
    \begin{aligned}\label{eq opt 2}
        && \underset{\substack{ \{\eta_{k}\}, t }}{\textrm{maximize}} & \quad t \\
        &&\textrm{subject to}&\quad 0\leq \eta_{k} \leq 1, \forall  k\in\mathcal{K}, \\
        &&& \quad \gamma^\text{ul}_{k} \geq t, \forall k\in \mathcal{K}.  
    \end{aligned}
\end{equation}
Recasting the problem in epigraph form simplifies the structure by replacing the non-smooth min operator with a scalar threshold variable $t$ and a set of SINR constraints. However, the resulting problem still involves fractional expressions within the SINR terms, making it challenging to solve efficiently. A conventional approach involves iteratively updating the fairness threshold $t$ and solving a sequence of linear programs to determine feasible power allocations. While this method can yield optimal solutions, it requires repeated re-solving and convergence checks, leading to significant computational overhead, especially in large-scale networks with many users and constraints. To overcome the inefficiency of iterative methods, the following subsection proposes a one-shot reformulation of problem \eqref{eq opt 2} into a QUBO form, enabling direct solution via QA.

Since \eqref{eq:R_ul_sinr}
is a strictly increasing function of $\gamma_k^{\text{ul}}$, maximizing the minimum uplink spectral efficiency is equivalent, in terms of the optimal power-control policy, to maximizing the minimum SINR. Therefore, in the following QUBO formulation, $t$ denotes the minimum SINR threshold. The corresponding max-min spectral efficiency is obtained after optimization as $\log_2(1+t)$. We first determine an upper bound $t_{\max}$ for the feasible SINR threshold. This can be done by evaluating each UE's individual peak SINR under full power while turning off all other users, i.e., $\eta_{k'}=0$ for $k'\neq k$. Hence, the SINR search interval is bounded by $t\in[0,t_{\max}]$. To make the problem suitable for binary encoding, we discretize both the SINR threshold $t$ and the power-control coefficients $\{\eta_k\}_{k=1}^{K}$ into finite levels:
\begin{itemize}
    \item Let $t\in\{t_1,t_2,\ldots,t_L\}$ be the set of $L$ candidate SINR thresholds sampled from $[0,t_{\max}]$.
    \item Let each power coefficient $\eta_k$ take values in a set $\mathcal{E}=\{e_1,\ldots,e_P\}\subset[0,1]$ of $P$ discrete levels.
\end{itemize}

We introduce binary indicator variables $z_{k,p}\in\{0,1\}$ for power selection such that
\begin{equation}
    \eta_k=\sum\nolimits_{p=1}^{P} z_{k,p}e_p,\qquad
    \sum\nolimits_{p=1}^{P} z_{k,p}=1,\quad \forall k\in\mathcal{K}.
\end{equation}
Similarly, let $v_{\ell}\in\{0,1\}$ indicate whether the SINR threshold $t_\ell$ is selected, with
\begin{equation}
    t=\sum\nolimits_{\ell=1}^{L}v_{\ell}t_\ell,\qquad
    \sum\nolimits_{\ell=1}^{L}v_{\ell}=1.
\end{equation}
For a fixed threshold level $t_\ell$, the SINR constraint $\gamma_k^{\text{ul}}\geq t_\ell$ can be equivalently rewritten as
$g_{k,\ell}(\{z_{k,p}\})\geq 0,\quad \forall k\in\mathcal{K}$, where $g_{k,\ell}$ is expressed as in \eqref{SINR-reform-true-qubo}. 
\begin{figure*}[t]
\vspace{-0.5cm}
\begin{equation}\label{SINR-reform-true-qubo}
\begin{aligned}
g_{k,\ell}(\{z_{k,p}\})
=&\;
\rho_u \sum\nolimits_{p=1}^{P} z_{k,p} e_p
\left(\sum\nolimits_{m=1}^{M} v_{mk}\right)^2 -
t_\ell
\Bigg(
\rho_u
\sum\nolimits_{\substack{k'=1\\k'\neq k}}^{K}
\sum\nolimits_{p=1}^{P}
z_{k',p}e_p
\left(
\sum\nolimits_{m=1}^{M}
v_{mk}
\frac{\beta_{mk'}}{\beta_{mk}}
\vert \boldsymbol{\phi}_{k'}^{H}\boldsymbol{\phi}_{k}\vert 
\right)^2
\\
&\qquad\quad
+
\rho_u
\sum\nolimits_{k'=1}^{K}
\sum\nolimits_{p=1}^{P}
z_{k',p}e_p
\sum\nolimits_{m=1}^{M}v_{mk}\beta_{mk'}
+
\sum\nolimits_{m=1}^{M}v_{mk}
\Bigg),
\quad \forall k\in\mathcal{K},\;\forall \ell\in\{1,\ldots,L\}.
\end{aligned}
\end{equation}
\noindent\rule{\textwidth}{0.5pt}
\end{figure*}
The binary epigraph problem can then be written as
\begin{subequations}\label{eq opt binary true qubo}
\begin{align}
    \underset{\{z_{k,p}\},\{v_\ell\}}{\textrm{maximize}}
    &\quad
    \sum\nolimits_{\ell=1}^{L}v_\ell t_\ell
    \\
    \textrm{subject to}
    &\quad
    \sum\nolimits_{p=1}^{P}z_{k,p}=1,\quad \forall k\in\mathcal{K},
    \\
    &\quad
    \sum\nolimits_{\ell=1}^{L}v_\ell=1,
    \\
    &\quad
    g_{k,\ell}(\{z_{k,p}\})
    +\Gamma_{k,\ell}(1-v_\ell)
    \geq 0, \quad \forall k\in\mathcal{K}, \nonumber\\ 
    & \quad\quad\quad\quad\quad\quad\quad\quad\quad\;\forall \ell\in\{1,\ldots,L\}.
\end{align}
\end{subequations}
Here, $\Gamma_{k,\ell}$ is a sufficiently large deactivation constant. It is chosen such that the constraint is automatically satisfied when $v_\ell=0$. A valid choice is any
\begin{equation}
    \Gamma_{k,\ell}
    \geq
    \left[
    -\min_{\{z_{k,p}\}\in\mathcal{Z}}
    g_{k,\ell}(\{z_{k,p}\})
    \right]_+,
\end{equation}
where $[x]_+=\max\{x,0\}$ and
\begin{equation}
    \mathcal{Z}
    =
    \left\{
    \{z_{k,p}\}:
    z_{k,p}\in\{0,1\},
    \sum_{p=1}^{P}z_{k,p}=1,\;\forall k\in\mathcal{K}
    \right\}.
\end{equation}
Thus, when $v_\ell=1$, the original SINR constraint $g_{k,\ell}(\{z_{k,p}\})\geq 0$ is enforced. When $v_\ell=0$, the same constraint is relaxed by the deactivation term $\Gamma_{k,\ell}$. To convert the inequality constraints into an unconstrained QUBO objective, we introduce binary slack variables $\psi_{k,\ell,r}\in\{0,1\}$ and define $s_{k,\ell} = \Delta_s \sum\nolimits_{r=0}^{R-1}2^r\psi_{k,\ell,r}$, where $\Delta_s>0$ controls the slack resolution and $R$ is the number of slack bits. The slack range should be selected such that
\begin{equation}
    \Delta_s(2^R-1)
    \geq
    \max_{\{z_{k,p}\}\in\mathcal{Z}}
    \left[
    g_{k,\ell}(\{z_{k,p}\})+\Gamma_{k,\ell}
    \right],
    \quad \forall k,\ell.
\end{equation}
{\color{black}
It should be noted that the binary slack expansion represents only the
discrete set
\begin{equation}
\mathcal{S}
=
\left\{
0,\Delta_s,2\Delta_s,\ldots,(2^R-1)\Delta_s
\right\}.
\end{equation}
Hence, condition~(30) ensures that the required residual range is covered,
whereas $\Delta_s$ determines the finite resolution of the slack
representation. Specifically, for any feasible residual
\begin{equation}
q_{k,\ell}
=
g_{k,\ell}(\{z_{k,p}\})
+
\Gamma_{k,\ell}(1-v_\ell)
\geq 0
\end{equation}
lying within this range, the closest representable slack value differs from
$q_{k,\ell}$ by at most $\Delta_s/2$. Therefore, the QUBO formulation below
should be interpreted as a quantized penalty representation of problem~(27);
the slack-representation accuracy improves as $\Delta_s$ is reduced. Given a selected slack resolution $\Delta_s$, the minimum number of slack bits required to cover the residual range can be chosen as
\begin{equation}
R=\left\lceil\log_2\left(\frac{r_{\max}}{\Delta_s}+1\right)\right\rceil,
\end{equation}
where $r_{\max}=\max_{k,\ell}\max_{\{z_{k,p}\}\in\mathcal{Z}}\left[g_{k,\ell}(\{z_{k,p}\})+\Gamma_{k,\ell}\right]$. Accordingly, the transformations up to the discretized problem are algebraically exact, while the approximation with respect to the original continuous formulation arises from the finite discretization of $\eta_k$ and $t$ and from the finite slack resolution $\Delta_s$.}
The resulting QUBO formulation is then given as in \eqref{eq:true_qubo}.
\begin{figure*}[t]
    \begin{equation}\label{eq:true_qubo}
    \begin{aligned}
    \min_{\{z_{k,p}\},\{v_\ell\},\{\psi_{k,\ell,r}\}}
    \quad
    & -\sum\nolimits_{\ell=1}^{L}v_\ell t_\ell + \lambda_1\sum\nolimits_{k=1}^{K}\left(\sum\nolimits_{p=1}^{P}z_{k,p}-1\right)^2 +\lambda_2\left(\sum\nolimits_{\ell=1}^{L}v_\ell-1\right)^2\\
    &+\lambda_3\sum\nolimits_{k=1}^{K}\sum\nolimits_{\ell=1}^{L}
    \Bigg(g_{k,\ell}(\{z_{k,p}\}) +\Gamma_{k,\ell}(1-v_\ell)-\Delta_s\sum\nolimits_{r=0}^{R-1}2^r\psi_{k,\ell,r}\Bigg)^2.
    \end{aligned}
    \end{equation}
    \noindent\rule{\textwidth}{0.5pt}
\end{figure*}
The first term maximizes the selected SINR threshold by minimizing its negative value. The second and third terms enforce the one-hot power and threshold selections, respectively. The fourth term enforces the SINR feasibility constraints through binary slack variables.

The formulation in \eqref{eq:true_qubo} is a true QUBO because each penalized expression is affine in the binary variables $\{z_{k,p}\}$, $\{v_\ell\}$, and $\{\psi_{k,\ell,r}\}$. After expansion and using $x^2=x$ for binary variables, the objective contains only constant, linear, and pairwise quadratic terms. Therefore, it can be expanded into the standard QUBO form $\min_{\mathbf{x}\in\{0,1\}^{N_b}}
\mathbf{x}^{T}\mathbf{Q}\mathbf{x}$, where $\mathbf{x}$ stacks all binary variables and $N_b=KP+L+KLR$. Here, $KLR$ is the number of slack variables when $R$ slack bits are assigned to each pair $(k,\ell)$. Applying the QCI diagnostic to problem~(22), the original formulation contains non-QUBO-native elements, including the $\min_k(\cdot)$ operator, the logarithmic mapping $\log_2(1+\gamma_k^{\mathrm{ul}})$, the fractional SINR expression $\gamma_k^{\mathrm{ul}}$, and the continuous power-control variables $\eta_k$, yielding $\mathrm{QCI}_{\mathrm{raw}}=0$. After the epigraph transformation, one-hot discretization, big-M deactivation, and binary slack encoding, \eqref{eq:true_qubo} contains only binary linear and pairwise-quadratic terms, so $\mathrm{QCI}_{\mathrm{post}}=1$. Table~4 distinguishes the parameters of the explicit QUBO encoding from the settings of the hybrid CQM implementation. For the explicit slack-based QUBO, the logical-variable count is $N_b = KP + L + KLR$. With $P=L=100$, this gives $11{,}100$ variables for $K=10$ and $R=10$, and $48{,}100$ variables for $K=40$ and $R=11$, including $10{,}000$ and $44{,}000$ slack variables, respectively. These are analytical counts for the explicit QUBO formulation, rather than variable counts of the submitted CQM or measurements of physical-qubit usage. Hence, $\mathrm{QCI}_{\mathrm{post}}=1$ does not imply a small or easily embeddable QUBO; QCI should be interpreted together with the encoded problem size, interaction density, and the requirements of the selected solver.



\subsection{Simulation results}

\begin{table}[t]
\centering
\caption{System parameters for the simulation.}
\renewcommand{\arraystretch}{1.5}
\label{table:simulation_parameters}
\resizebox{\linewidth}{!}{
\begin{tabular}{|l|c||l|c|} 
\hline
\textbf{Parameter} & \textbf{Value} &\textbf{Parameter} & \textbf{Value} \\ \hline\hline
Carrier frequency & 1.9 GHz & Bandwidth & 20 MHz \\ \hline
Noise & 9 dB & AP antenna height & 15 (m) \\ \hline
User antenna height & 1.65 (m) & Shadowing standard deviation & 8 dB \\ \hline
Service area diameter & 1000 m & Number of APs & (30, 100) \\ \hline
Number of users & (10, 40) & Pilot sequence length & 20 \\ \hline
\end{tabular}
}
\end{table}

{\color{black}
\begin{table}[!t]
\centering
\caption{\textcolor{black}{Formulation, encoding, and QPU settings used in the case-study experiments.}}
\label{tab:qpu_settings}
\renewcommand{\arraystretch}{1.15}

\begin{tabularx}{\linewidth}{@{}p{0.43\linewidth}X@{}}
\toprule
Parameter & Value \\
\midrule
\multicolumn{2}{@{}l}{\textit{Discretization}} \\
Power levels $P$ / threshold levels $L$
    & $100/100$ \\
\midrule
\multicolumn{2}{@{}l}{\textit{Explicit QUBO encoding}} \\
Slack bits $R$ / slack step $\Delta_s$
    & $(10,0.2)$ for $K=10$;
      $(11,0.5)$ for $K=40$ \\
Penalty coefficients $(\lambda_1,\lambda_2,\lambda_3)$
    & $(50,50,10^8)$ \\
\midrule
\multicolumn{2}{@{}l}{\textit{Hybrid CQM implementation}} \\
Sampler
    & \texttt{LeapHybrid}\allowbreak\texttt{CQMSampler} \\
Constraint handling
    & Explicit one-hot and SINR constraints \\
Objective
    & Negative selected SINR threshold \\
Explicit penalty/slack terms
    & Not used in the submitted CQM \\
SINR constraint scaling
    & $10^8$ \\
Requested time limit
    & At least $150$~s; adjusted to the
      solver-reported minimum \\
Solution selection
    & Lowest-objective feasible returned sample \\
QPU timing field
    & \texttt{qpu\_access\_time} \\
\bottomrule
\end{tabularx}
\par\smallskip
\begin{minipage}{\linewidth}
\footnotesize
\emph{Note:} The slack and penalty parameters describe
the explicit QUBO encoding and are not settings of
the submitted CQM. The implementation scales the
SINR constraints by $10^8$; this factor is not an
objective penalty. When the solver minimum
$T_{\min}$ is available, the requested time limit
is $\max\{150,\lfloor T_{\min}\rfloor+10\}$ seconds;
otherwise, it is $150$ seconds.
\end{minipage}
\end{table}
\begin{figure}[t]
    \centering
    \subfloat[{\scriptsize 30 APs, 10 Users}]{\includegraphics[width=0.50\columnwidth]{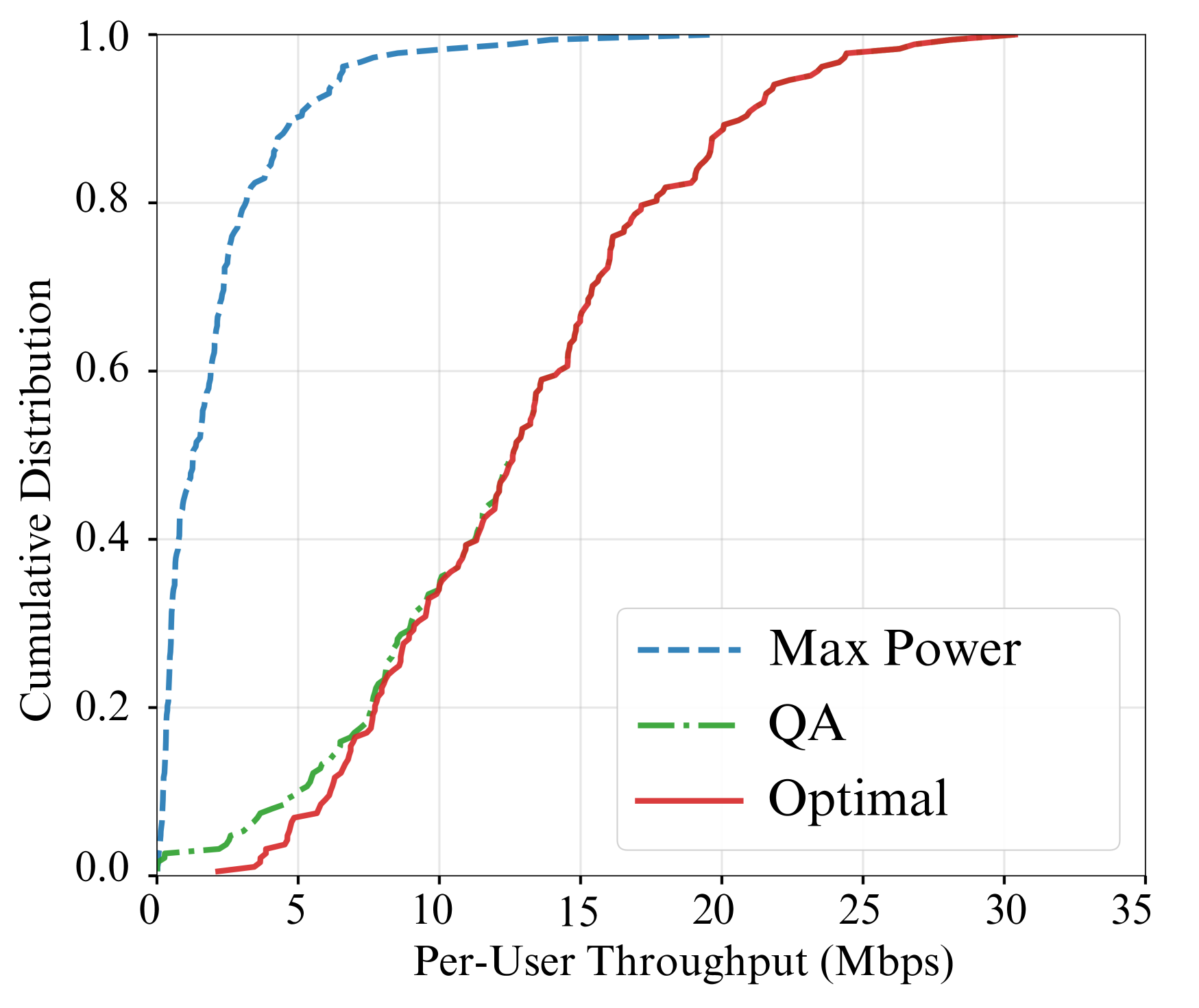}}
    \subfloat[{\scriptsize 100 APs, 40 Users}]{\includegraphics[width=0.50\columnwidth]{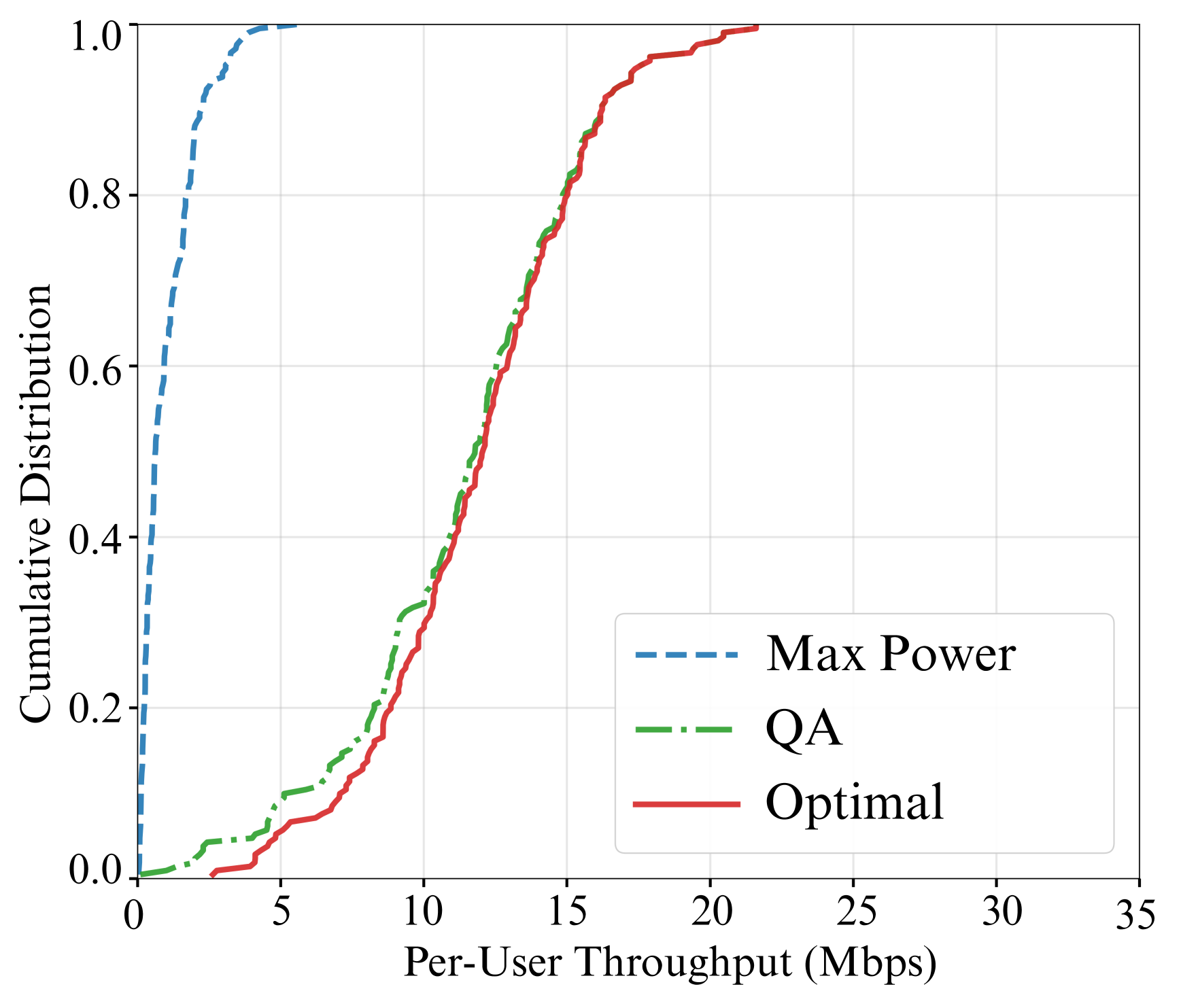}}
    \caption{CDF of the uplink ergodic throughput per user.}
    \vspace{-0.5cm}
    \label{cdf_result}
\end{figure}

To validate the proposed QA-based optimization, we simulate a cell-free massive MIMO network under two different scales: a small configuration with $30$ APs and $10$ UEs, and a large configuration with $100$ APs and $40$ UEs. The evaluation is conducted in a one kilometer service area with the system configuration summarized in Table~\ref{table:simulation_parameters}. The QA method is benchmarked against two references: (i) the Optimal solution derived using the iterative bisection algorithm in \cite{2017_TWC_Hien_CfmMIMO}, and (ii) a Max Power baseline, where all users transmit with full power. The numerical experiments were performed using D-Wave's \texttt{LeapHybridCQMSampler}. The implementation retains the one-hot and SINR constraints explicitly in a constrained quadratic model (CQM), rather than introducing the penalty and slack terms of the explicit QUBO encoding. In the figures and tables, the label QA refers to this hybrid quantum-classical CQM implementation. All experiments for the classical baseline methods were performed on a workstation running Windows~$11$, equipped with $32$~GB of RAM and an AMD Ryzen~$5$~$7600$ $6$-Core Processor. The system performance is reported in terms of the cumulative distribution function (CDF) and box plots of the average ergodic throughput.

The CDF results in Fig.~\ref{cdf_result} show that the throughput distribution achieved by the QA-based method closely follows that of the bisection-based reference in both network scales. For the tested instances, the hybrid CQM implementation achieves a throughput distribution close to that of the bisection-based reference. The Max Power baseline, by contrast, yields a substantially lower throughput distribution, since uncontrolled full-power transmission increases inter-user interference. Fig.~\ref{rate_distribution} provides complementary statistical insight. The median and interquartile range of the QA-based solution are close to optimum, while remaining consistently higher than those of the Max Power baseline. 
\begin{figure}[t]
    \centering
    \subfloat[{\scriptsize 30 APs, 10 Users}]{\includegraphics[width=0.50\columnwidth]{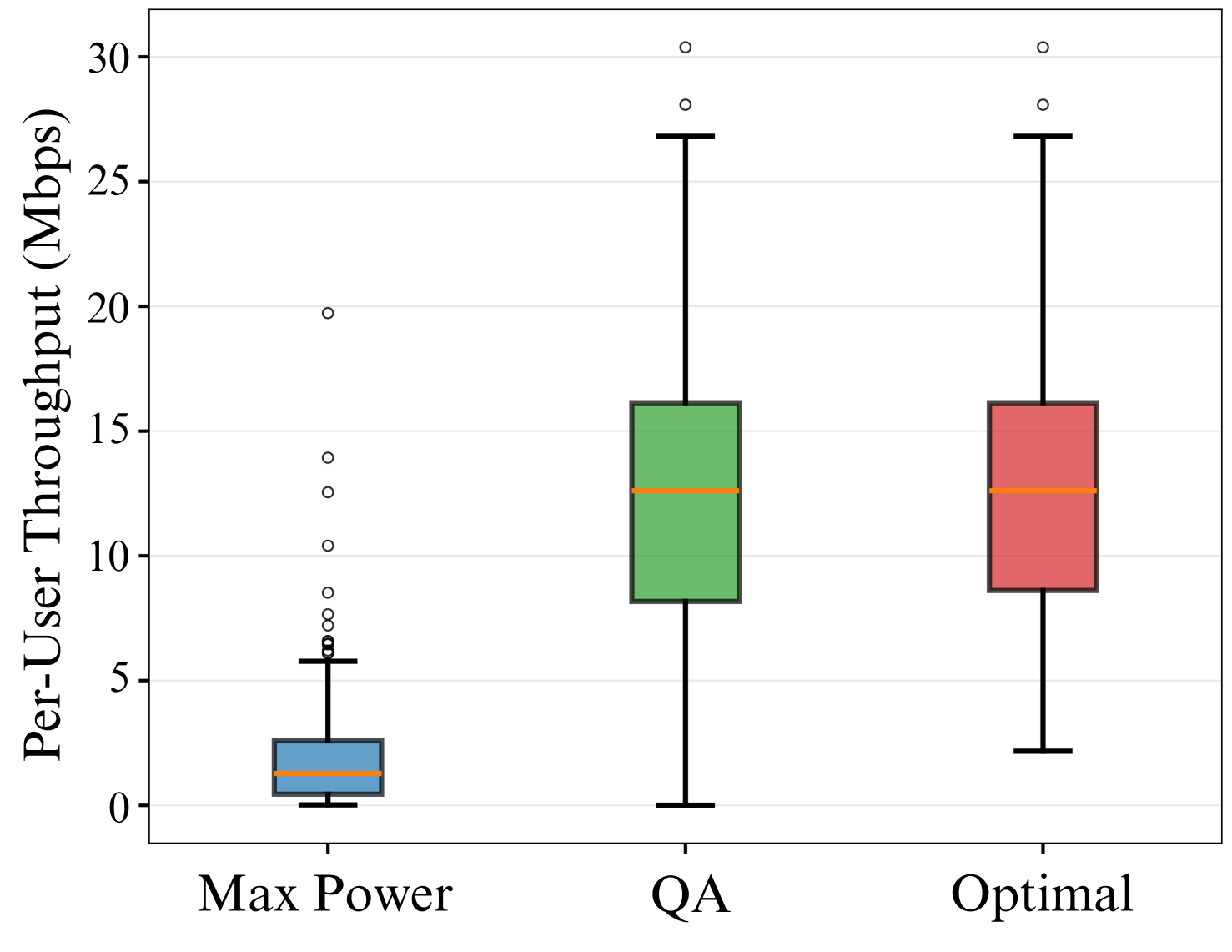}}
    \subfloat[{\scriptsize 100 APs, 40 Users}]{\includegraphics[width=0.50\columnwidth]{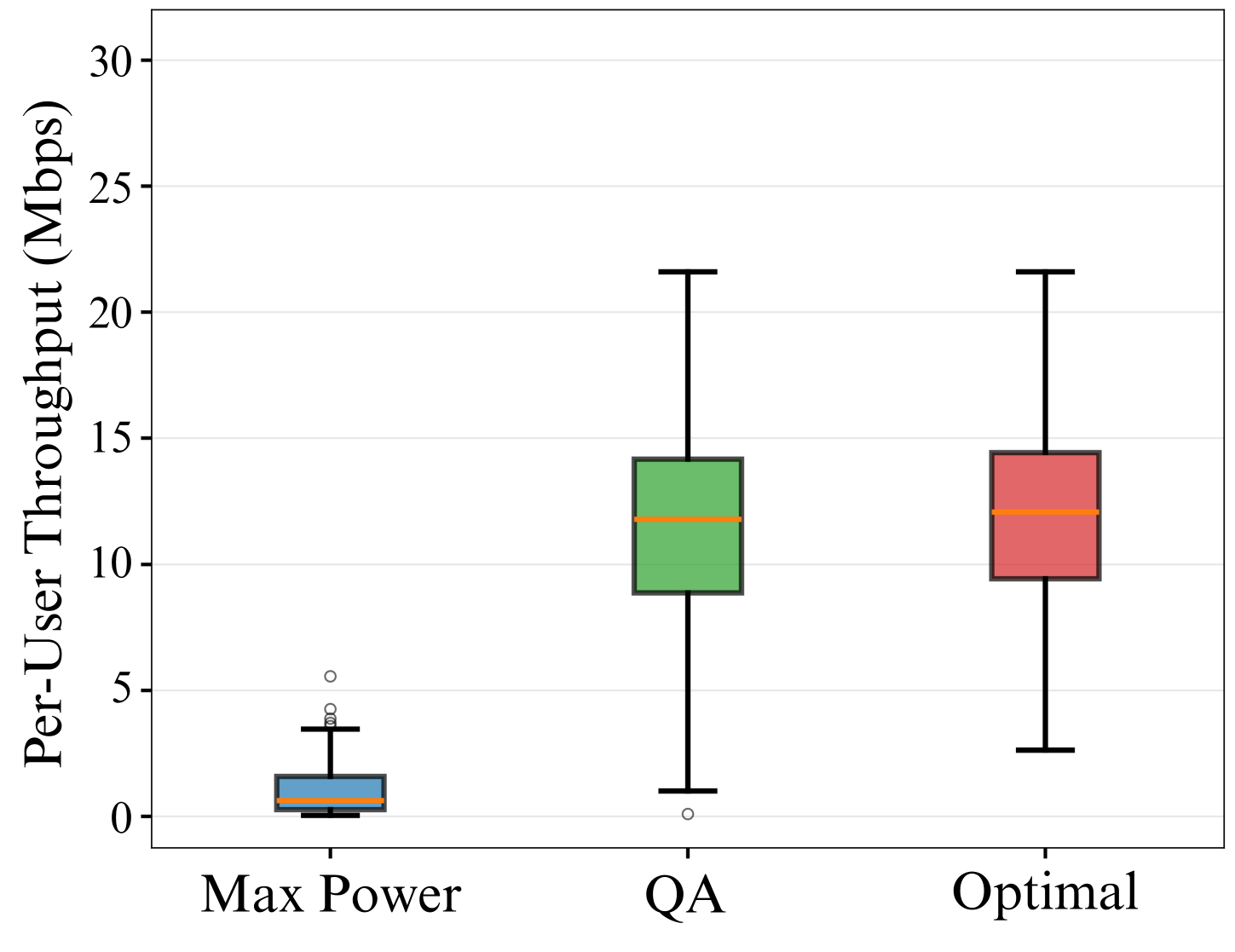}}
     \caption{Comparison of the average ergodic throughput.}
    \vspace{-0.35cm}
    \label{rate_distribution}
\end{figure}
\begin{table}[!t]
\centering
\caption{\textcolor{black}{Average minimum user spectral efficiency, corresponding to the max-min objective value in problem (22). The gap denotes the relative loss of QA versus the optimal solution.}}
\label{tab:minrate_r2}
\scriptsize
\setlength{\tabcolsep}{3.3pt}
\color{black}{
\begin{tabular}{lccccc c}
\toprule
\multirow{2}{*}{Configuration} &
\multicolumn{5}{c}{$\bar{R}_{\min}$ (bit/s/Hz)} &
\multirow{2}{*}{Gap (\%)} \\
\cmidrule(lr){2-6}
& Max Power & SA & TS & QA & Optimal & \\
\midrule
30 APs-10 UEs  & 0.18 & 0.76 & 0.81 & 0.86 & 0.89 & 3.37 \\
100 APs-40 UEs & 0.11 & 0.45 & 0.63 & 0.66 & 0.70 & 5.71 \\
\bottomrule
\end{tabular}
}
\end{table}
\textcolor{black}{
To directly evaluate the max–min fairness objective in~\eqref{eq opt 1}, Table~\ref{tab:minrate_r2} reports the average minimum-user spectral efficiency. For the configuration with $30$ APs and $10$ UEs, the QA-based method achieves $0.86$ bit/s/Hz, compared with $0.89$ bit/s/Hz for the optimal solution obtained via bisection, corresponding to a relative gap of only $3.37\%$. For the larger configuration with $100$ APs and $40$ UEs, QA achieves $0.66$ bit/s/Hz, compared with the optimal value of $0.70$ bit/s/Hz, resulting in a relative gap of $5.71\%$. We also benchmark the proposed formulation against two classical
metaheuristics, namely simulated annealing (SA) and tabu search (TS). Both yield a lower minimum-user spectral efficiency than QA result, with TS attaining $0.81$ and $0.63$ bit/s/Hz and SA attaining $0.76$ and $0.45$ bit/s/Hz in the two configurations, the degradation being more pronounced for SA in the larger network. The Max Power baseline performs poorly in both the small- and large-scale network configurations, achieving average minimum-user spectral efficiencies of only $0.18$ and $0.11$ bit/s/Hz, respectively.
}

\begin{table}[t]
\centering
\caption{Classical-method runtimes and solver-reported QPU access times of the hybrid CQM method.}
\renewcommand{\arraystretch}{1.25}
\label{tab:runtime_compar}
\color{black}{
\begin{tabular}{|c|c|c|c|}
\hline
          & 30 APs--10 UEs & 60 APs--20 UEs & 100 APs--40 UEs \\ \hline
Bisection & 1.078 s        & 3.438 s        & 8.568 s         \\ \hline
SA & 3.078 s        & 4.379 s        & 5.568 s         \\ \hline 
Tabu & 3.758 s        & 5.348 s        & 6.068 s         \\ \hline
QA        & 0.034 s        & 0.034 s        & 0.034 s         \\ \hline
\end{tabular}
}
\begin{minipage}{0.95\linewidth}
\vspace{1mm}
\footnotesize
\emph{Note:} The QA row reports the \texttt{qpu\_access\_time} returned by the hybrid CQM solver, converted from microseconds to seconds. The corresponding workflow-level interpretation is provided in Table~\ref{tab:complexity_compar}.
\end{minipage}
\vspace{-0.4cm}
\end{table}
Table~\ref{tab:runtime_compar} reports the runtimes of the classical methods and the solver-reported QPU access time of the hybrid CQM implementation for three network configurations. The bisection runtime increases from $1.078$~s to $3.438$~s and $8.568$~s as the network size grows. This trend is expected because the bisection method repeatedly solves a feasibility subproblem while refining the SINR threshold. The runtime of the two metaheuristics grows more slowly than that of the bisection method, from $3.078$~s to $5.568$~s for SA and from $3.758$~s to $6.068$~s for Tabu. In contrast, the QA row remains approximately constant for the tested cases because it reports the QPU execution time under fixed annealing and sampling settings.

\subsection{Complexity analysis}
Table~\ref{tab:complexity_compar} summarizes the dominant complexity terms of the classical reference methods and an illustrative direct-QPU workflow for the explicit QUBO formulation. The latter is distinct from the hybrid CQM implementation used in the numerical experiments. For the bisection method, let $\epsilon_t$ be the bisection tolerance and let $[0,t_{\max}]$ be the SINR search interval. The number of bisection iterations is $I_{\mathrm{bis}}
= \left\lceil
\log_2\left(\frac{t_{\max}}{\epsilon_t}\right)
\right\rceil$ . For each fixed SINR threshold, the feasibility problem has $K$ power-control variables and $\mathcal{O}(K)$ feasibility constraints. However, each user constraint depends on the interference contributions from the other users, so forming the coupled constraint coefficients requires $\mathcal{O}(K^2)$ operations after coefficient preprocessing. Let $C_{\mathrm{LP}}(K)$ denote the cost of solving one LP feasibility subproblem. The per-iteration complexity of the bisection method is  $C_{\mathrm{bis}}^{\mathrm{iter}} =
\mathcal{O}\!\left(K^2+C_{\mathrm{LP}}(K)\right)$. For a dense interior-point LP implementation, the LP solution cost can be approximated as $C_{\mathrm{LP}}(K) \approx
\mathcal{O}\!\left(I_{\mathrm{LP}}K^3\right)$, where $I_{\mathrm{LP}}$ denotes the number of interior-point iterations. Therefore, including the $\mathcal{O}(MK^2)$ coefficient preprocessing cost, the overall bisection complexity becomes
\begin{equation}
\begin{aligned}
C_{\mathrm{bis}}
&=
\mathcal{O}\!\left(
MK^2+
I_{\mathrm{bis}}
C_{\mathrm{bis}}^{\mathrm{iter}}
\right)\\
&=
\mathcal{O}\!\left(
MK^2+
I_{\mathrm{bis}}
\left[
K^2+C_{\mathrm{LP}}(K)
\right]
\right).
\end{aligned}
\end{equation}
Under the dense interior-point approximation, this becomes
\begin{equation}
\begin{aligned}
C_{\mathrm{bis}}
&=
\mathcal{O}\!\left(
MK^2+
I_{\mathrm{bis}}
\left[
K^2+I_{\mathrm{LP}}K^3
\right]
\right)\\
&\approx
\mathcal{O}\!\left(
MK^2+I_{\mathrm{bis}}I_{\mathrm{LP}}K^3
\right).
\end{aligned}
\end{equation}
This explicitly quantifies the product between the number of bisection iterations and the per-iteration computational cost. Hence, the increasing bisection runtime in Table~\ref{tab:runtime_compar} is consistent with the growth of both the LP dimension and the inter-user coupling terms as $K$ increases. For the Max Power baseline, no optimization problem is solved. Its computational burden is dominated by evaluating the SINR and throughput expressions for all users. In a dense evaluation, this requires accumulating inter-user interference terms over $M$ APs and $K$ UEs, leading to the complexity $C_{\mathrm{MP}}=\mathcal{O}(MK^2)$. 

For the QA workflow, the number of logical binary variables is $N_b = KP + L + KLR$, where $P$ is the number of discrete power levels, $L$ is the number of discrete SINR-threshold levels, and $R$ is the number of slack bits assigned to each pair $(k,\ell)$. The QUBO expansion cost is governed by the one-hot constraints and the SINR-feasibility penalty terms. A conservative upper-bound scaling is
\begin{equation}
C_{\mathrm{exp}}
=
\mathcal{O}\!\left(
KP^2+L^2+KL(KP+R+1)^2
\right).
\end{equation}
Thus, the QUBO construction complexity can be written as
\begin{equation}
C_{\mathrm{QUBO}}
=
\mathcal{O}\!\left(MK^2+C_{\mathrm{exp}}\right),
\end{equation}
where the $\mathcal{O}(MK^2)$ term accounts for coefficient computation and $C_{\mathrm{exp}}$ accounts for constructing the binary quadratic model. For an illustrative direct-QPU implementation of
this QUBO, a component-wise time model is
\begin{equation}
T_{\mathrm{QA}}
=
T_{\mathrm{coef}}
+
T_{\mathrm{QUBO}}
+
T_{\mathrm{emb}}
+
T_{\mathrm{QPU}}
+
T_{\mathrm{post}},
\end{equation}
where $T_{\mathrm{coef}}$ is the coefficient-computation time, $T_{\mathrm{QUBO}}$ is the QUBO construction time, $T_{\mathrm{emb}}$ is the minor-embedding time, $T_{\mathrm{QPU}}$ is the QPU execution time, and $T_{\mathrm{post}}$ includes decoding, chain repair, feasibility checking, and post-processing.
\begin{table*}[t]
\centering
\caption{Complexity of the classical reference methods and an illustrative direct-QPU time model for the explicit QUBO formulation.}
\label{tab:complexity_compar}
\footnotesize
\renewcommand{\arraystretch}{1.35}
\begin{tabular}{|p{0.12\textwidth}|p{0.13\textwidth}|p{0.3\textwidth}|p{0.35\textwidth}|}
\hline
\textbf{Method}
& \textbf{Role}
& \textbf{Iterations / samples}
& \textbf{Dominant complexity / time model}
\\ \hline

Max Power
&
Baseline
&
One SINR/rate evaluation
&
$C_{\mathrm{MP}}=\mathcal{O}(MK^2)$
\\ \hline

Bisection + LP
&
Reference method
&
$I_{\mathrm{bis}}=
\left\lceil
\log_2\left(t_{\max}/\epsilon_t\right)
\right\rceil$
&
$C_{\mathrm{bis}}
=
\mathcal{O}\!\left(MK^2+
I_{\mathrm{bis}}C_{\mathrm{bis}}^{\mathrm{iter}}\right)$
\\ \hline

QA/QUBO
&
Proposed method
&
One QUBO instance, sampled $N_{\mathrm{reads}}$ times
&
$T_{\mathrm{QA}}
=
T_{\mathrm{coef}}
+
T_{\mathrm{QUBO}}
+
T_{\mathrm{emb}}
+
T_{\mathrm{QPU}}
+
T_{\mathrm{post}}$
\\ \hline

\end{tabular}
\vspace{-0.2cm}
\end{table*}
The complexity comparison in Table~\ref{tab:complexity_compar} complements the runtime measurements in Table~\ref{tab:runtime_compar}. The Max Power baseline has the lowest computational burden because it only evaluates the resulting SINR/rate values without optimizing the power coefficients. The bisection-based reference has a higher cost because the LP feasibility problem must be solved repeatedly over $I_{\mathrm{bis}}$ threshold-refinement steps. The QA/QUBO method avoids this outer bisection loop, but its workflow-level cost depends on the QUBO size, embedding, QPU sampling, and post-processing. Therefore, the nearly constant QA values in Table~\ref{tab:runtime_compar} should be interpreted as favorable solver-side QPU latency for the tested QUBO instances, while the overall scalability behavior is reflected by the QA/QUBO time model in Table~\ref{tab:complexity_compar}.

\section{Challenges and Future Directions}

QA offers a promising computational paradigm for solving complex wireless resource allocation problems. However, its practical deployment in NG-Networks remains constrained by several hardware and algorithmic limitations. This section highlights key challenges and outlines the corresponding research directions to overcome these limitations.

\subsection{Challenge 1: Scalability of Hardware and Wireless Problem Dimensionality}
\textbf{Challenge:} Current QA hardware (e.g., D-Wave Advantage2) offers a limited qubit count with sparse interconnectivity. Embedding dense wireless optimization problems into such topologies requires minor-embedding~\cite{minor_embedding}, introducing significant overhead. Furthermore, realistic wireless tasks involve multi-domain resource allocation (e.g., power, spectrum, access points), leading to high-dimensional QUBO formulations that exceed current hardware capacity.  

\textbf{Future Direction:} Scalable embeddings may be achieved through advanced decomposition strategies, graph partitioning~\cite{ushijima2017graph}, and sparse approximations~\cite{ayanzadeh2020ensemble} that exploit wireless problem structure (e.g., interference sparsity). Hybrid quantum-classical frameworks~\cite{qa_for_optimize} should be further explored to delegate dense subproblems to classical solvers, while reserving combinatorial cores for QA. 

\subsection{Challenge 2: Noise and Reliability of QA}
\textbf{Challenge:} QA systems remain highly sensitive to noise. Analog control errors and calibration imprecision distort the problem Hamiltonian, while thermal excitations reduce ground-state fidelity even in cryogenic temperatures. Such distortions undermine reliability in wireless contexts, where performance metrics (e.g., SINR, power levels) are sensitive to small deviations.  

\textbf{Future Direction:} Robust encodings~\cite{pearson2019analog}, noise-aware formulations~\cite{jeong2025embedding}, and Hamiltonian regularization~\cite{roch2023effect} can mitigate analog distortions. Adaptive annealing schedules~\cite{finvzgar2024designing}, and ensemble-based filtering~\cite{ayanzadeh2020ensemble} should be refined, potentially guided by reinforcement learning. In the long term, error-corrected QA systems are expected to fundamentally enhance reliability.  

\subsection{Challenge 3: QUBO Representation of Wireless Problems}

\textbf{Challenge:} Wireless resource-allocation problems often include continuous or integer variables, while QA hardware natively supports binary variables. These variables must therefore be quantized or binary-encoded, which increases the logical-variable count, physical-qubit usage, and embedding overhead, creating a tradeoff between discretization accuracy and QUBO size. Another difficulty arises from the fact that wireless performance metrics, such as spectral efficiency and rate fairness, often depend on logarithmic SINR functions, which are nonlinear and non-quadratic. Their QUBO conversion therefore requires surrogate approximations, such as polynomial or piecewise-linear models, which may alter the energy landscape and reduce solution fidelity.

\textbf{Future Direction:} Promising directions include developing adaptive discretization schemes~\cite{arai2023effectiveness} that balance accuracy and qubit efficiency, together with problem-specific encodings that reduce embedding overhead. Hybrid quantum-classical methods are also promising, with classical preprocessing handling nonlinear terms and QA targeting the core combinatorial structure. Advanced approximations, including higher-order polynomial mappings~\cite{norimoto2023quantum} and quantum-native reformulations~\cite{chancellor2019domain}, may further improve the fidelity of QUBO-based wireless models.

\subsection{Challenge 4: Integration with Dynamic Wireless Environments}
\textbf{Challenge:} Wireless systems operate in highly dynamic environments with time-varying channels and traffic patterns. Current QA workflows tend to require complete re-embedding for each scenario~\cite{jeong2025quantum}, limiting real-time adaptability. Moreover, existing QA solvers are designed for generic optimization and lack integration with wireless system simulators, constraining cross-layer evaluation.  

\textbf{Future Direction:} Incremental embedding methods~\cite{okada2019improving}, warm-started annealing~\cite{pelofske2023initial}, and embedding reuse~\cite{lobe2021embedding} can enable near real-time re-optimization for latency-sensitive applications. Developing wireless-aware quantum solvers with interfaces to platforms will further enable direct evaluation. Hardware-in-the-loop frameworks~\cite{chanda2023architecture} could support live benchmarking under realistic radio conditions.  

\subsection{Challenge 5: Standardization and Benchmarking}
\textbf{Challenge:} The absence of standardized QA formulations and wireless benchmarks hinders reproducibility, comparative evaluation, and industrial adoption. Currently, it is difficult to assess QA performance against classical baselines in a consistent manner.  

\textbf{Future Direction:} Broader community is desired for establishing standardized datasets and libraries (e.g., QA-Wireless Suite) covering canonical wireless tasks such as power control, user association, and scheduling. Unified evaluation protocols and performance metrics, such as embedding overhead, energy gap, and fidelity are essential to benchmark QA against classical and hybrid solvers. 

\section{Conclusions}
Wireless-oriented QUBO modeling guidelines for applying QA to resource allocation in NG networks were presented. 
A rudimentary terrestrial-aerial communication network scheduling example was used
to illustrate the end-to-end mapping from an interference-coupled SINR objective to a QUBO/Ising representation and the corresponding scheduling decision. This example can be applied naturally to UAV-assisted, aircraft-assisted, and satellite-assisted links in SAGIN systems. We reviewed the foundations of QA, discussed the BQM/DQM/CQM solver ecosystem, and summarized key modeling steps, including variable encoding, penalty-based constraint handling, embedding considerations, solution decoding, and feasibility verification. To support formulation-level assessment, we introduced the QUBO Compatibility Index (QCI) as a lightweight diagnostic, while emphasizing that it should be interpreted together with post-encoding and hardware-aware metrics. A worked case study on max-min uplink power control in cell-free massive MIMO illustrated how a wireless fairness problem can be discretized and converted into a true QUBO formulation. The numerical results illustrate the performance of a hybrid CQM implementation on the discretized power-control instances considered, while the explicit QUBO encoding provides a formulation-level description of the penalty-based approach. Finally, we discussed key challenges and future directions, including hardware limitations, noise sensitivity, QUBO representation fidelity, dynamic network adaptation, and standardized benchmarking for QA-assisted wireless optimization.
\bibliographystyle{IEEEtran}
\bibliography{reference}


\end{document}